\documentclass[12pt]{article}
\usepackage{amsmath,amsfonts,amssymb,bbm}
\usepackage{graphics}
\usepackage{amsfonts}
\usepackage{amsmath}

\usepackage[latin1]{inputenc}
\usepackage[pdftex]{graphicx}

\usepackage{color}
\usepackage[american]{babel}
\usepackage{cite} 

\usepackage[top=2.5cm, bottom=3cm, left=2.5cm, right=2.5cm]{geometry}

\makeatletter \@addtoreset{equation}{section}

\newcommand{\be}{\begin{equation}}
\newcommand{\ee}{\end{equation}}
\newcommand{\bea}{\begin{eqnarray}}
\newcommand{\eea}{\end{eqnarray}}
\newcommand{\bmat}{\begin{bmatrix}}
\newcommand{\emat}{\end{bmatrix}}

\newcommand{\bbibitem}[1]{\bibitem{#1}\marginpar{#1}}

\def\Label#1{\label{#1}%
  \smash{\hbox to0pt{\raise1ex\hbox{\tiny[#1]}\hss}}}
\def\noLabels{\let\Label=\label}
\def\nobbibitem{\let\bbibitem=\bibitem}

\newcommand{\p}{\partial}

\begin{document}
%
\renewcommand{\thefootnote}{\fnsymbol{footnote}}
\vspace{0truecm}
\thispagestyle{empty}

\hfill
\vspace{1.5truecm}
\begin{center}
{\fontsize{21}{18} \bf Dynamical Love Numbers\\ [7 pt] 
for Kerr Black Holes}\\[14pt]
\end{center}

\vspace{.15truecm}

\begin{center}
{\fontsize{13}{18}\selectfont
Malcolm Perry,$^{\rm a,b,c}$\footnote{\texttt{malcolm@damtp.cam.ac.uk}} Maria J.~Rodriguez,$^{\rm d,e,f}$\footnote{\texttt{majo.rodriguez.b@gmail.com}} }\\[4.5pt]
\end{center}
\vspace{.4truecm}

\begin{scriptsize}
 \centerline{{\it ${}^{\rm a}$ Department of Physics and Astronomy, Queen Mary University of London, Mile End Road, London E1 4NS, UK}} 
 
  \vspace{.05cm}
 
 \centerline{{\it ${}^{\rm b}$ DAMTP, Centre for Mathematical Sciences, University of Cambridge, Wilberforce Road, Cambridge CB3 0WA, UK}}
 
  \vspace{.05cm}
 
 \centerline{{\it ${}^{\rm c}$ Trinity College, Cambridge, CB2 1TQ, UK}}
 
   \vspace{.05cm}

 \centerline{{\it ${}^{\rm d}$Instituto de Fisica Teorica, Universidad Autonoma de Madrid, 13-15 Calle Nicolas Cabrera, 28049 Madrid, Spain}}

  \vspace{.05cm}

\centerline{{\it ${}^{\rm e}$Department of Physics, Utah State University, 4415 Old Main Hill Road, UT 84322, USA}}

  \vspace{.05cm}

 \centerline{{\it ${}^{\rm f}$ Black Hole Initiative, Harvard University, 20 Garden Street, Cambridge MA 02138, USA}}
\end{scriptsize}
 \vspace{.25cm}

\vspace{.3cm}
\begin{abstract}
\noindent

While {\it static} Love number vanish identically for Kerr black holes, we show that the corresponding {\it dynamical} tidal coefficients are generically non-zero and exhibit logarithmic behavior. The computational method employs a related but simpler scheme consistent with CFT descriptions, low-frequency regimes and post-Newtonian results. These coefficients are illustrated with a numerical examples.

\end{abstract}

\newpage

\setcounter{tocdepth}{2}
\tableofcontents
\newpage
\renewcommand*{\thefootnote}{\arabic{footnote}}
\setcounter{footnote}{0}

\section{Introduction}

Tidal deformation is a significant phenomenon observed in various astronomical scenarios, including the merger of binary black holes, binary stellar systems (referenced, for instance, \cite{Gurlebeck:2015xpa,Chatziioannou:2012gq}), sources within our own Solar System (famously, for instance, the Sun, Earth, and Moon system or Jupiter's moon Io) and exoplanets (e.g. \cite{Barros_2022}). To assess and study the tidal forces exerted by some companion on a central star or planet, the idea of tidal Love numbers were introduced by Augustus Love in 1909. These numbers quantify the magnitude of an induced multipole moment in response to external field acting on the central object by its companion. These objectives can be examined, as Love himself did, within the context of linear response.

By now, we have become accustomed to the idea that rotating Kerr black holes suffer no tidal distortion in  response to external static gravitational perturbations. Following a contentious series of publications, a consensus has been reached regarding the vanishing of the static tidal response coefficients, colloquially termed the Love numbers \cite{Binnington:2009bb,Fang:2005qq,Damour:2009vw,Kol:2011vg,Hui:2020xxx,Chia:2020yla,Goldberger:2020fot, Charalambous:2021mea}. These static Love numbers characterize time-independent tidal deformations. However, as we argue here, the size and shape of the event horizon of a black hole will change in response to external time-dependent gravitational perturbations. The essence of these time-dependent perturbations can be developed in a way that manifestly preserves hidden approximate near-horizon $SL(2,R) \times SL(2,R)$ symmetries. The purpose of the present paper is to carry this out for the case of time-dependent tidal deformations by establishing the dynamical Love numbers for Kerr black holes.

The gravitational tidal coefficients, $ k_{\ell m}$, describe the tidal response of an elastic object, for example a star, a planet or a black hole. In Section \ref{sec:Setup} we give a reference to the precise definitions of these coefficients and explain our conventions.
The tidal coefficients $ k_{\ell m}$ at a frequency $\omega$ can be complex but are more conveniently divided up into their real and imaginary parts. Thus
\bea\label{Love def}
k_{\ell m}(\omega)= \kappa_{\ell m}(\omega) + i \, \nu_{\ell m}(\omega) \, 
\eea
where ${\ell}$ and $m$ label  the conventional  spherical harmonics.
Dissipative effects are encoded in imaginary parts  $\nu_{\ell m}$, whereas the real part  captures the tidal deformation Love numbers $\kappa_{\ell m}$.
The distinction between $\kappa_ {\ell m} $ and $\nu_{\ell m}$  has a simple physical analogy in electromagnetism. There, the real part captures conservative effects such as refraction, while the imaginary part of the electric susceptibility gives rise to the dissipation. 
Potential uncertainties in the tidal coefficient arising from the extension of Newtonian results
to a general relativistic setting has been  addressed by utilizing an analytic continuations of the GR solutions into higher dimensions. This approach, as detailed in the work of Kol and Smolkin \cite{Kol:2011vg}, essentially corresponds to treating the parameter $\ell$ as a non-integer value.
A different strategy to address these uncertainties involves utilizing the framework defined within the point-particle effective field theory (EFT) for binary inspirals \cite{Goldberger:2004jt,Porto:2016pyg}.

Remarkably, the real part of the static Love numbers $\kappa_{\ell m}(\omega=0)$, which determine the response to time-independent external fields, are found to vanish in four-dimensional Einstein theory for both Schwarzschild and  Kerr black holes. By contrast, there is a non-zero dissipative part 
$\nu_{\ell m}(\omega=0)$ for Kerr black holes when $\omega=0$ as a result of frame dragging.
For a brief review of static Love numbers for Kerr black holes see Section \ref{sec:Review}. Whilst the tidal coefficients for Kerr black holes $k_{\ell m}(\omega=0)$ are well known, the dynamical tidal response and corresponding Love number $k_{\ell m}(\omega\ne0)$ are not well understood, and the primary goal of this work is to introduce them and supply them with a CFT interpretation.\\

The dynamical tidal response of black holes is captured by the frequency-dependent Love numbers which quantify the induced multipole moments the object acquires when an external field varies with time in the object's rest frame. Further, in the PPN formalism, these characteristic tidal numbers manifest themselves at various different Post-Newtonian (PN) orders in the phase of emitted gravitational waves when the object is part of a binary system. In particular, the manifestation of tidal deformation in a body, $\kappa_{\ell m}$, becomes evident at the 5PN order in the phase of a binary waveform \cite{Vines:2011ud,Bini:2012gu}. On the other hand, the onset of tidal dissipation in a rotating body, encoded in $\nu_{\ell m}$, is observed for the first time at the 2.5PN order for Kerr and 4PN order for Schwarzschild \cite{Poisson:1994yf,Alvi:2001mx,Poisson:2004cw,Porto:2007qi}. An accurate measurement of these tidal effects would not only provide valuable insights into the characteristics of known objects like neutron stars or black holes, but also could potentially indicate the existence of new types of compact objects
like hypothetical quark stars.

Finding the dynamical Love numbers has been shown to suffer from a drawback: the Teukolsky's equation at finite frequency can only be solved numerically or for certain limiting regimes. Some of the proposals that have been put forward to determine these coefficients rely on the so called Kerr Effective Geometries (KEG) \cite{Lowe:2011aa,Charalambous:2021kcz,Hui:2022vbh,Castro:2010fd}. KEG are characterized by their property of approximating the near zone environment of the black hole while preserving Kerr's structure on the horizon and at the same time realizing certain approximate {\it hidden} conformal symmetries \footnote{In contrast to a Killing vector symmetry of the black hole geometry.} of the Klein-Gordon wave equation in the KEGs background. Two different instances of KEGs were recently proposed as explanations for the vanishing of the static Love numbers. The first example invokes an $SO(4,2)$ hidden symmetry, as suggested in \cite{Hui:2022vbh}. This is what the authors referred to as ``effective near zone geometry" for Kerr black holes and  \cite{Charalambous:2022rre} called Starobisnky near-zone metric. A second example proposes that there exists an effective metric with an $SL(2,R) \times U(1)$ hidden symmetry, as demonstrated in the reference \cite{Charalambous:2021kcz} and referred to as the Love near zone metric in \cite{Charalambous:2022rre}. While such symmetry arguments for the vanishing of Love numbers for Kerr are quite promising, the proposed KEGs do not seem to capture the black hole tidal response at non-zero frequency $\omega$ as has been shown in \cite{Charalambous:2022rre}.
Clearly, one wishes to obtain the dynamical Love numbers for Kerr black holes that are compatible with the known results for the small-spin limit at order $\mathcal{O}(\Omega^2)$ and for the low-order frequency perturbation of $\mathcal{O}(\omega \Omega, \omega^2)$, where here $\Omega$ is the angular velocity of the Kerr black hole. This raises the question: what is the simplest effective description of the Kerr black hole that captures these small-spin and low-order frequency dynamical Love numbers?
 
Building upon previous work on the subject \cite{Castro:2010fd}, we employ an approach that is consistent with a CFT description of the Kerr black hole, and manifestly preserves the hidden $SL(2,R) \times SL(2,R)$ symmetries. We present a systematic calculation of dynamical Love numbers for Kerr black holes using a perturbative approach and show that the tidal response coefficients exhibit logarithmic contributions, which are universal and independent of any renormalization scheme. The computational method we employ is related to a simpler scheme consistent with CFT descriptions \cite{Castro:2010fd}, low-frequency regimes obtained in the form of a series over hypergeometric functions \cite{Mano:1996vt} and PN results \cite{Poisson:2020vap}. In addition to providing a new understanding of the frequency-dependent response of Kerr black holes, our results provide new insights on dynamical tidal deformation of Schwarzschild black holes.

In this work, we accordingly compute the dynamical Love numbers for Kerr black holes. We introduce the notation and briefly review the methods for calculating static Love number for Kerr black holes in Section \ref{sec:Review}. In Section \ref{sec:Dynamical} we find that all frequency dependent contributions to the Love numbers $\kappa_{\ell m}(\omega)$. Our approach focuses on the KEG in \cite{Castro:2010fd}. We argue that frequency-dependent tidal response coefficients display logarithmic behavior and in turn are universal and independent of the renormalization scheme for both, static Schwarzschild and Kerr black holes. We analyze the low-order frequency perturbation in Section \ref{sec:Low} and, show that tidal Love coefficients in the near-black hole region with hidden $SL(2,R) \times SL(2,R)$  symmetry capture the frequency dependent behavior of a Kerr black hole. In Section \ref{sec:Effective} we present a comparison with the non-universal dynamical tidal coefficients previously derived in alternative KEG backgrounds developed in \cite{Charalambous:2021kcz, Hui:2022vbh}. Possible CFTs descriptions are included in Section \ref{sec:CFT}. Section \ref{sec:Discussion} includes a discussion of our results.


\section{Equations for Tidal Response Coefficients}
\subsection{Teukolsky Equation}
\label{sec:Setup}
In this Section we present the equations needed to calculate the Kerr black hole tidal response coefficients and we also discuss some of the important technical details necessary to find
the correct expressions.

The gravitational tidal coefficients, $ k_{\ell m}$, describe the tidal response of a rigid object e.g. star, planet or black hole. For a non-rotating spherical body of mass $M$ and
equilibrium radius $r_s$ \footnote{ In our context $r_s$ will be the black hole's Schwarzschild radius $r_s=2M $. Other choices will lead to trivial rescalings of response coefficients by a constant factor.}, one can show that the total Newtonian potential in spherical coordinates produced by an external static perturbation takes the form
\bea\label{newtonian}
\Phi=-\frac{M}{r}+\frac{(\ell-2)!}{\ell!} \sum_{\ell=2} \sum_{m=-\ell}^{\ell} Y_{\ell m} \,\mathcal{E}_{\ell m} \, r ^{\ell} \left[ 1+ k_{\ell m} \left(\frac{r}{r_s}\right)^{-2 \ell -1} \right]\,,
\eea
where $\mathcal{E}_{\ell m}$ are the angular harmonic coefficients and $Y_{lm}$ the spherical harmonics. The perturbation is labelled in terms of the scalar harmonics and the leading contribution in \eqref{newtonian} corresponds to the body's internal multipole moments.

In the General Relativity (GR), it is convenient to look at the metric perturbation for the description of the tidal deformation of the rigid object by writing
\bea
g_{\mu\nu}=\eta_{\mu\nu} + \, h_{\mu\nu}\,
\eea
with $\eta_{\mu\nu}$ being the flat space Minkowski space background.
The linearized Einstein equations for a small perturbation $h_{\mu\nu}$ around flat Minkowski space is
\bea
\Box  \, \bar{h}_{\mu\nu} = -16 \pi  \,  T_{\mu\nu}\,,
\eea
where the energy-momentum tensor is $ T_{\mu\nu}= (2/\sqrt{-g} ) (\delta I_{matter}/\delta \, g^{\mu\nu})$,  $\bar{h}_{\mu\nu} = { h}_{\mu\nu} - \frac{1}{2}\, h \,\eta_{\mu\nu}$ and the transvere gauge choice  $\nabla_{\mu} \bar{h}^{\mu\nu}=0 $ has been made.

In the case of black holes in 4D, by direct analogy with the Newtonian case, the temporal metric component perturbation $h_{tt}$ for Schwarzschild or Kerr can be written in the long-distance limit as
\bea\label{perturbation}
2\,h_{tt}=\frac{M}{r} -\frac{(\ell-2)!}{\ell!} \sum_{\ell=2} \sum_{m=-\ell}^{\ell}Y_{\ell m} \,\mathcal{E}_{\ell m} \, r ^{\ell} \left[ \left(1+\mathcal{O}(r^{-1}) \right)+ k_{\ell m} \left(\frac{r}{r_s}\right)^{-2 \ell -1} \left(1+\mathcal{O}(r^{-1}) \right)\right]
\eea
By comparison with the Newtonian gravitational, asymptotically the expression \eqref{perturbation} offers a useful way to extract tidal response coefficients from a Newtonian gravitational potential $\Phi$ generated by an external source in GR since
in the weak-field limit
\bea
\Phi=-\frac{1}{2}\,h_{tt}\,.
\eea

Therefore, at first glance, this provides us with a practical prescription, such that the tidal response coefficients for Kerr black holes can be computed by using the 
Teukolsky equation (found by  Teukolsky on \cite{Teukolsky:1973ha} for the linearization of the Einstein equations in Kerr). For Kerr black holes, this reduces to a simple generalization of the vacuum Klein-Gordon (KG) equation to deal with fields of spin-$s$. The detailed relation between $\Phi_s$ and the field perturbations can be found in \cite{Teukolsky:1972}. An important observation is that the tidal response coefficients can be extracted directly from the solutions of these equations for all integer spin fields to all orders in the frequency, including both static ($\omega=0$) and dynamical ($\omega\ne 0$) responses. 

Let us start by considering the Teukolsky equation for a massless $s$-spin perturbation in the Kerr spacetime background in Boyer-Lindquist coordinates. The Teukolsky equation completely separates into a trivial time dependent piece with frequency $\omega$, a trivial dependence on the azimuthal quantum number $m$ and
more complicated polar angle and radial functions. 
\
A general solution of the Teukolsky equation can be then be written as 
\be\label{eq:spin}
\Phi_s(t,r,\theta,\phi)= e^{-i \omega t+i m\phi} R_s(r) S_s(\theta) \ , \quad \text{with}\quad \omega \in \mathbb{C} \quad \text{and} \quad m\in \mathbb{Z}\ .
\ee
The function $S_s, R_s$ satisfy ODEs with appropriate boundary conditions that ensure $\Phi_s$ is a well behaved smooth function (see details here below). Formally $s$ can assume any integral value, and are related to scalar ($s=0$), electromagnetic ($s=\pm1$), and gravitational ($s=\pm 2$) perturbations around the Kerr black hole with mass $M$ and reduced spin parameter $a\equiv J/M$. 

For $s=0,\pm 1,\pm 2$, we define $R_s(r) := \Delta^{-s/2}\hat{R}_s(r)$ where  the function $\Delta=(r-r_+)(r-r_-)$. Let $r_\pm=M\pm\sqrt{M^2-a^2}$ then $r_+$ and $r_-$ label respectively the outer and inner horizons of the Kerr black hole. The radial equation for a particle of spin $s$ can be written as  
\bea\label{eq:radial}
&\!\!\!\!\!\!\!\!\!\!&\Bigg[\p_r \Delta \p_r +  \frac{\left(2M\omega r_+ - \frac{i}{2} s(r_+ - r_-) - a m\right)^2}{(r-r_+)(r_+ - r_-)}  - \frac{\left(2M\omega r_- + \frac{i}{2} s (r_+ - r_-) - a m\right)^2}{(r-r_-)(r_+ - r_-)}    \qquad\qquad\qquad\qquad  \\
&\!\!\!\!\!\!\!\!\!\!& \quad\qquad\qquad\qquad\qquad\qquad + \omega^2 r^2 + 2(M\omega+is)\omega r + 2M \,\omega\,(2M\,\omega - is) - s^2 - K_{\ell,s} \Bigg] \hat{R}_s(r) = 0 \ .  \nonumber
\eea 
Note that the radial ODE is not real when $s\neq 0$. In the radial equation $K_{\ell,s}$ 
is the separation constant and is determined by the  eigenvalue of the spin-weighted spheroidal equation
for the polar angle dependence
\bea\label{eq:spheroidal}
\left[\frac{1}{\sin\theta} \p_\theta \left( \sin\theta \p_\theta \right) + (a\omega\cos\theta-s)^2 - \frac{(m+s\cos\theta)^2}{\sin^2\theta} - s^2 + K_{\ell,s} \right] S_s(\theta) = 0  \ .
\eea
The value of angular momentum $\ell$ and hence $K_{\ell,s}$ is found from the normalizability of the spheroidal eigenfunctions.

Both equations \eqref{eq:radial} and \eqref{eq:spheroidal} are the confluent Heun type, each with two regular singular points and a irregular singular point of Poincar\'e rank one.

The solutions of the spheroidal equation for the Kerr black hole with arbitrary frequency are difficult to obtain in general, and require numerical or sophisticated analytical methods. 
There is no closed-form solution to this equation for the rotating black hole with arbitrary frequency. One common approach is to use approximate methods for certain regimes of the parameters. In the slowly rotating small frequency regime $a\omega \ll 1$ we expect $K_{\ell,s}=\sum_{n\ge0} a_n (a \omega)^{n}$. For the use of the next sections, we quote the first orders of the coefficient\footnote{It is also worth mentioning that Mathematica has a built in function for computing $K_{\ell,s}$ called SpheroidalEigenvalue.}
\bea\label{separationconst}
K_{\ell,s}=(\ell-s)(\ell+s+1)+s- a \omega \,\frac{2 m (s^2+\ell (\ell+1))}{\ell(\ell+1)}+ \mathcal{O}(a^2\omega^2)\,.
\eea

Finally, a comment on the radial equation $\eqref{eq:radial}$ is in order. The solution to the radial ODE again has no closed-form expression, but can obtained as a series of hypergeometric functions \cite{Mano:1996vt}. The ODE has two regular singular points  at the horizons $r=r_{\pm}$ and the irregular singular point is at $r=\infty$. Only in the static limit, all three singular points in the ODE are regular singular points. This implies that the static problem can be solved in terms of  hypergeometric functions, (see Section \ref{sec:Review} for more details). 

For non-zero values of the frequency $\omega \ne 0$, various inequivalent approaches incorporating approximate symmetry considerations have been proposed to describe accurately certain near-zone regimes
in such a way that the irregular singular point at $r=\infty$ is replaced by a regular singular point.
Consequently, solutions to the ODE can be expressed as representations of the hidden $SL(2,R)$ symmetries leading to solutions in terms of hypergeometric functions. There exists some uncertainty in delineating the near zone to Kerr black holes due to the variability in the $\omega$ regimes; distinct KEG can characterize the Kerr near zone differently and so producing different differential equations within 
presumably valid within their own 
specific regimes. We provide reasoning that suggests the Kerr/CFT inspired approximation in \cite{Castro:2010fd}, in contrast to KEG proposals in \cite{Charalambous:2021kcz,Hui:2022vbh}, is {\it the} framework capable of producing the dynamical Love numbers for Kerr black holes. For further insights, see Section \ref{sec:Dynamical}.

\subsection{Boundary Conditions} 
The radial functions must meet the following ingoing boundary conditions at the horizon
\bea\label{eq:asympt}
 \hat{R}_s(r)= const \times (r-r_+)^{-i\alpha_+}\,,\, \text{with} \,\, \alpha_+ >0 \qquad \text{as} \qquad r\rightarrow r_+\,.
\eea
where we defined the coefficient
\bea
\alpha_+ \equiv \frac{(\omega- m\,\Omega)}{4\pi\,T_+}\pm\frac{i s}{2}\,
\eea
The black hole's angular velocity is given by $\Omega=a/(2 M r_+)$ and the Hawking temperature (or equivalently the surface gravity $\kappa=2\pi T_+$) by $T_+=(r_+ - r_-)/ (8\pi M r_+)$. In general, for the full dynamical problem $\omega\ne 0$, as described in detail by the authors in \cite{Castro:2013kea}, the solutions that describe outgoing or ingoing waves around the event horizon $r=r_+$ respectively are of the form $R_{out}= (r-r_+)^ {i\alpha_+} (1+O(r-r_+))$ and $R_{in}= (r-r_+)^ {-i\alpha_+} (1+O(r-r_+))$  when $\omega\,\alpha_+>0$. These roles are reversed when $\omega \, \alpha_+<0$. Note that for the static $\omega=0$ limit, $\alpha_+<0$ but the role is not reversed. This coincides with boundary condition guarantees regularity of the radial solution at the event horizon that the energy momentum flux flows strictly into the black hole \cite{Teukolsky:1972}.

To match the source at the boundary, the constant in \eqref{eq:asympt} can be chosen such that 
\bea
 \hat{R}_s(r)/r^{\ell}\rightarrow 1 \qquad  \text{as} \qquad r\rightarrow \infty.
\eea
After establishing the appropriate boundary conditions for $\Phi_s$, a practical approach to determining the tidal response coefficients involves analyzing the behavior at spatial infinity. This entails reading off the coefficients in front of $r^{-\ell-1}$ by performing a Taylor expansion of the solution at the spacetime boundary. It is crucial to emphasize that this technique might potentially result in gauge-independent ambiguities, making it challenging to accurately identify source and response terms.\footnote {Indeed, this factor was the root cause of the confusion that prompted multiple authors to assert the existence of non-zero Love numbers for Kerr black holes \cite{LeTiec:2020spy}}. The key ingredient of our procedure is the analytic continuation of the relevant static response solutions to non-integer values of the orbital multipole number  $\ell \in \mathbb{R}$. This will allow us to extract the response coefficients in an invariant way as justified in e.g. \cite{Binnington:2009bb,Kol:2011vg}.

\section{A Review of Static Love Numbers for Kerr}
\label{sec:Review}

The purpose of this section is twofold: to establish the notation used throughout the rest of the paper and to ensure that the paper is comprehensible without external references.  We will now derive the static tidal response coefficients, also known as Love numbers, specifically for a Kerr black hole. Our focus will be on the examination of the static response, \eqref{eq:radial} with $\omega=0$, for bosonic (integer) spin perturbations. The radial equation is now 
\bea\label{radial}
&\!\!\!\!\!\!\!\!\!\!&\Bigg[\p_r \Delta \p_r +  \frac{\left( - \frac{i}{2} s(r_+ - r_-) - a m\right)^2}{(r-r_+)(r_+ - r_-)}  - \frac{\left(\frac{i}{2} s (r_+ - r_-) - a m\right)^2}{(r-r_-)(r_+ - r_-)} - s^2 - K_{\ell,s} \Bigg] \hat{R}_s(r) = 0
\eea 
in which no approximations have been made.  In the static case, as shown in \cite{Teukolsky:1972}, the separation constant is $K_{\ell,s}=(\ell-s)(\ell+s+1)+s$. 

The first step in the analysis is to write the radial equation (\ref{radial}) as a hypergeometric differential equation
\bea\label{eqHyper}
z\,(1-z) \,\, \frac{d^2 w}{dz^2}+[\mathfrak{c}-(\mathfrak{a}+\mathfrak{b}+1)\,z] \, \,\frac{dw}{dz}-a b \,\, w=0\,
\eea
which has all three singular points  $z = 0,1,\infty$ being regular. Starting with (\ref{radial}) we proceed by making the following change of coordinates, and redefinitions of the functions
\bea\label{sols}
 r=r_++(r-r_-) \,z\,,\qquad \hat{R}_s(r)=(r-r_-)^p\,(r-r_+)^q\,w\,,
\eea
In the new coordinates the event horizon $r=r_+$ is located at $z=0$, the inner horizon $r=r_-$ at $z=\infty$ and $r=\infty$ at $z=1.$

After some algebra we choose the coefficients \footnote{For Schwarzschild, when the spin parameter vanishes $a=0$, the exponents become $p= q=- s/2$.}
%
%
%
%
\bea\label{pqvalues}
p = -  i\gamma m- \frac{s}{2} \,,\qquad q=  i\gamma m - \frac{s}{2}\,,
\eea
with $\gamma= a/(r_+-r_-)$, which provides a convenient way of ensuring the solutions $\hat{R}_s(r)$ be smooth at the outer black hole horizon. In turn, the values (\ref{eqHyper}) yield
\bea\label{coefficients}
\mathfrak{a}= 1+\ell-s \,,\qquad \mathfrak{b}= 1+\ell+ 2 i m \gamma\,,\qquad  \mathfrak{c}=1-s+2 i m  \gamma\,,
\eea
%
%
Having identified the relevant coefficients \eqref{coefficients} in the differential equations we can now analyze its solutions in the neighborhood of the singular points (see Abramowitz \& Stegun \cite{AB1964}). The full solution can be written in the following way
\bea
 \hat{R}_s(z)&=& (1-z)^p\,z^q\, (1-z)^\mathfrak{a}(c_1 F[\mathfrak{a},\mathfrak{b},\mathfrak{c};z]+c_2 \,z^{1-c} \,F[\mathfrak{a}-\mathfrak{c}+1,\mathfrak{b}-\mathfrak{c}+1,2-\mathfrak{c};z])\,,\nonumber
\eea
where $c_1,c_2$ are constants that will be fixed with by the choice of boundary conditions and $F$ stands for the Gauss hypergeometric function. Employing the values in \eqref{pqvalues} the solution becomes
\bea
 \hat{R}_s(z)&=&c_1
 \,(1-z)^{-  i\gamma m- \frac{s}{2} } \,z^{ i\gamma m - \frac{s}{2}}  \,F[1+ \ell- s,1+\ell+2 i m \gamma,1-s+2 i m \gamma;z]\\
 &&+ \,c_2   \,(1-z)^{-  i\gamma m- \frac{s}{2} } \,z^{-( i\gamma m - \frac{s}{2})}\, \,F[1+\ell+s ,1+\ell-2 i m \gamma,1+s-2 i m  \gamma;z]  \,)\nonumber\,.
\eea
We are working at the classical level, so the radial function $\hat{R}_s$ has to be regular across the horizon, and thus one sets $c_2=0$. 

\begin{figure}
\centering
  \includegraphics[width=8cm, height=6cm]{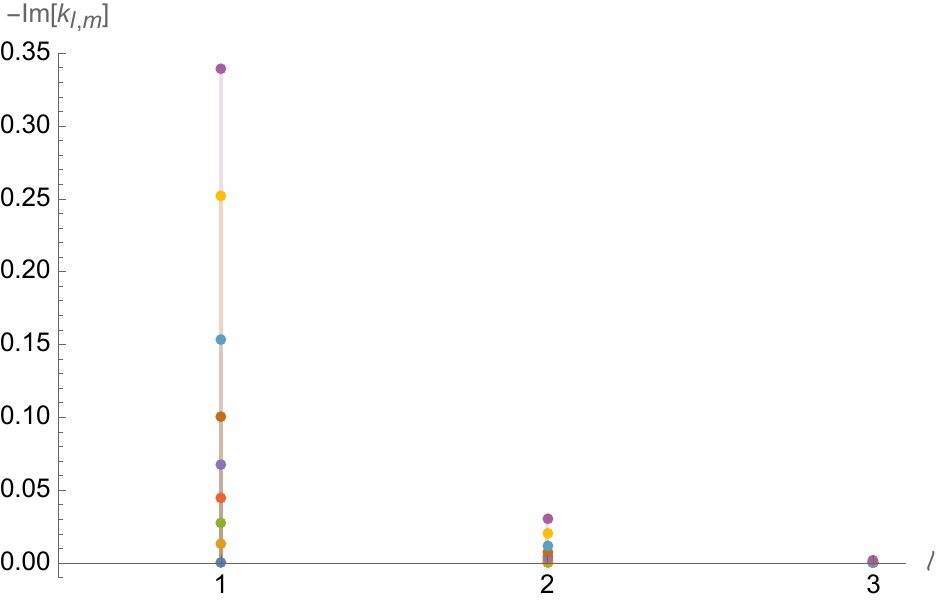}
  \caption{Visualization of the response coefficients $\lambda^{Kerr}_{\ell m}$ for Kerr black holes \eqref{KerrLove2}  (in units of $(r_s/(2M))^{1+2 \ell}$) as a function of the multipole moments $\ell$ for various values of the Kerr black hole spin $J= a M$. The real part of the coefficients vanish $Re(\lambda^{Kerr}_{\ell m})=0$, leading to vanishing static Love numbers. The non trivial dissipation coefficients, defined as the imaginary part of the response coefficients  $Im(\lambda^{Kerr}_{\ell m})= \nu_{lm}(\omega=0)$, are represented here. Choosing a fixed value of the multipole moment $l$, as the Kerr black hole spin increases, with rotational parameter $a=0.1,0.2,0.3,0.4,0.5,0.6,0.7,0.8$ (from {\it lightblue}  to {\it purple} or upwards), with fixed mass $M=1$, $s=0$ and $m=1$, the dissipation parameters becomes larger. The behavior changes as one compares the dissipation coefficients fixing all values but $\ell$; for increasing multipole value $\ell$ the dissipation decreases.}
  \label{fig1}
\end{figure}

One is now interested in the large $r$, $z\rightarrow 1$, behavior of the ingoing near-region solution. To achieve this, one uses the $z\rightarrow 1-z$ linear transformation formula for the hypergeometric function
\bea
F[\mathfrak{a},\mathfrak{b},\mathfrak{a}+\mathfrak{b}+1-\mathfrak{c};z]&=&\frac{\Gamma(\mathfrak{c})\Gamma(\mathfrak{c}-\mathfrak{a}-\mathfrak{b})}{\Gamma(\mathfrak{c}-\mathfrak{a})\Gamma(\mathfrak{c}-\mathfrak{b})}F[\mathfrak{a},\mathfrak{b},\mathfrak{a}-\mathfrak{b}-\mathfrak{c}+1;1-z]\nonumber\\
 &&+z^{\mathfrak{c}-\mathfrak{a}-\mathfrak{b}}\,\frac{\Gamma(\mathfrak{c})\Gamma(\mathfrak{a}+\mathfrak{b}-\mathfrak{c})}{\Gamma(\mathfrak{a})\Gamma(\mathfrak{b})} F[\mathfrak{c}-\mathfrak{a},\mathfrak{c}-\mathfrak{b},\mathfrak{c}-\mathfrak{a}-\mathfrak{b}+1;1-z]\,.\nonumber
\eea
To avoid ambiguities resulting from cancellations between body's response and the subleading source contributions, we consider an analytic continuation in which $\ell \rightarrow \mathbb{C}$. The large $r$ behavior of the ingoing wave solution is then given by
%
%
\bea
\hat{R}_s(r) \xrightarrow[r\rightarrow\infty] \,\tilde{c_1} \, r^{\ell} \left(1+ \left(\frac{r}{r_+-r_-}\right)^{-(1+2\ell)}\,\frac{\Gamma(-2\ell-1)\Gamma(1+\ell-s)\Gamma(1+\ell+2mi\gamma)}{\Gamma(- \ell-s)\Gamma(- \ell+2 m i \gamma)\Gamma(2\ell+1)} \right)\,,
\eea
which comparing with the expression in (\ref{perturbation}) and considering
\bea
\hat{R}_s(r) \xrightarrow[r\rightarrow\infty] \,\tilde{c_1} \, r^{\ell} \left(1+ \left(\frac{r}{r_+ +r_-}\right)^{-(1+2\ell)}\, k_{\ell m}\right)\,,
\eea
provides us with the following dimensionless response coefficients
\bea
k_{\ell m}(\omega=0)=  \frac{\Gamma(-2\ell -1)\, \Gamma(1+\ell -s)\, \Gamma(1+\ell+2mi\gamma)}{\Gamma(- \ell-s)\,\Gamma(- \ell+2 m i \gamma)\,\Gamma(2\ell+1)} \left( \frac{r_+-r_-}{r_++r_-}\right)^{(1+2\ell)}\,,
\eea
We can now set $\ell \in \mathbb{N}$. Using the properties of the Euler gamma function (see App. \ref{sec:gamma}) we find
\bea\label{KerrLove1}
k_{\ell m}(\omega=0)= \kappa_{\ell m}(\omega=0)+ i \,\nu_{\ell m}(\omega=0)\,,
\eea
where
\bea\label{KerrLove2} 
\kappa_{\ell m}(\omega=0)&=&0\,, \\
 \nu_{\ell m}(\omega=0)&=&(-1)^{s+1}m \gamma \frac{(\ell+s)! (\ell-s)!}{(2 \ell+1)! (2\ell)!}\left( \prod_{n=1}^{\ell} (n^2+ 4m^2\gamma^2)\right) \left( \frac{r_+-r_-}{r_++r_-}\right)^{(1+2\ell)}\,.
\eea
Interestingly, the relevant response coefficients are purely imaginary in this case. While $\kappa_{\ell m}(\omega=0)$ vanishes and defines the {\it static} Love number, $\nu_{\ell m}(\omega=0)$ is the tidal dissipative response coefficient. Figure \ref{fig1} portrays the dissipative coefficient of Kerr black holes. As depicted, when the black hole's angular momentum increases (for fixed values of $\ell$), the dissipative coefficients also increase. On the other hand, when the value of $\ell$ is increased, the dissipative coefficients $ \nu_{\ell m}(\omega=0)$ decrease monotonically. This trend is similar in higher dimensional black holes and black rings, as has been observed in previous studies \cite{Castro:2013lba}. Beyond the static response, the dissipative response coefficient has been shown not to vanish  $\nu_{\ell m}(\omega\ne0) \ne 0$ and exhibit a hidden symmetry \cite{Charalambous:2021mea,Hui:2021vcv}. The dynamical responses are only known in the small spin regime to orders $\mathcal{O}(\omega^2, \omega \Omega,\Omega^2)$. Next, we will compute the frequency dependent Love numbers for Kerr black holes $\kappa_{\ell m}(\omega)$ and show these do not vanish and capture the slow spin response coefficients for both Kerr black holes and the non-rotating static Schwarzschild black holes.

\section{Dynamical Love Numbers for Kerr} 
\label{sec:Dynamical}

In this Section  we compute the frequency-dependent extension of the tidal responses. The focus of this section will be the near-zone black hole regime analyzed in \cite{Castro:2010fd}. The proposal linked  finite temperature CFT correlators precisely with the Kerr scattering amplitudes when the wavelength of the scalar excitation is large compared to the radius of curvature. That is when
\bea\label{eq:condition}
\omega M \ll 1\,, \qquad \omega r \ll 1\,.
\eea
The solution space of the wave equation for the propagating field has  conformal symmetry even when the space on which the field propagates does not. In contrast with alternative frequency regimes linked to different KEGs \cite{Charalambous:2021mea,Hui:2021vcv}, there is technical advantage of the near zone approximation that we propose for the computation. The $SL(2,R) \times SL(2,R)$  regime in \cite{Castro:2010fd} provides a simple way of computing the frequency-dependent extensions for the dynamical Love numbers that were until now only known at first order for slowly spinning black holes. Our systematic calculations show that the dynamical Love numbers also receive corrections for Schwarzschild black holes already at order $\omega ^2$ in the frequency.

\subsection{Computation}

The wave equation \eqref{eq:radial} simplifies considerably for \eqref{eq:condition}. In this regime, as shown in \cite{Castro:2010fd}, it is possible to find a range of parameters where the order $\omega^2$ in the second line of the Teukolsky equation \eqref{eq:radial} can be neglected. Therefore, one possibility to study the dynamical response in Kerr black holes is to consider near region using the following equation
\begin{small}
\bea\label{radialOmega}
\Bigg[\p_r \Delta \p_r +  \frac{\left(2M\omega r_+ - \frac{i}{2} s(r_+ - r_-) - a m\right)^2}{(r-r_+)(r_+ - r_-)}  - \frac{\left(2M\omega r_- + \frac{i}{2} s (r_+ - r_-) - a m\right)^2}{(r-r_-)(r_+ - r_-)}  -  \hat{K}_{\ell, s} \Bigg] \hat{R}_s = 0
\eea 
\end{small}
\noindent where $\hat{K}_{\ell, s}=s^2 + K_{\ell ,s} $ is the separation constant. 

As far as the angular problem is concerned, the leading near zone approximation is chosen to coincide with the static approximation
\be\label{spheroidal}
\left[\frac{1}{\sin\theta} \p_\theta \left( \sin\theta \p_\theta \right) - \frac{(m+ s\cos\theta)^2}{\sin^2\theta}  + K_{\ell,s} \right] S_s(\theta) = 0  \ .
\ee
In the near region, the angular equation reduces to the Laplacian on the 2-sphere. The eigenfunctions are then given by the standard spin weighted spherical harmonics, yielding $K_{\ell,s}=(\ell-s)(\ell+s+1)+s$. The orbital number $\ell \ge |s|$ is always an integer in the leading near zone approximation.

In order to address the radial ODE, Eq. \eqref{radialOmega}, it is worth highlighting that the equation possesses three distinct regular singular points (located at $r=r_{\pm},\infty$) and so can be reduced to the  hypergeometric equation (\ref{eqHyper}). 
As before we use
\bea\label{sols1}
  z=\frac{r-r_+}{r-r_-}\,,\qquad \hat{R}_s(r)=(r-r_-)^p\,(r-r_+)^q\,w(r)\,,
\eea
but now  the coefficients are chosen to be 
\bea
 p =\frac{2 i M r_+ (\omega - m \Omega)}{r_+-r_-} -(1+\ell)\,,\qquad q= -\frac{2 i M r_+  (\omega - m \Omega)}{r_+-r_-} =- i \,\alpha_+\,,
 \eea
and the values in (\ref{eqHyper}) are then
\bea
&&\mathfrak{a}= 1+\ell-i\frac{4 M}{r_+-r_-} (M\omega -r_+ m \Omega)\,,\qquad \mathfrak{b}= 1+\ell- 2 i M \omega-s\,,\\
&&\mathfrak{c}=1-i\frac{4 M\,r_+}{r_+-r_-} (\omega - m \Omega)-s\,.
\eea
where
\bea
\mathfrak{c}-\mathfrak{a}-\mathfrak{b}= -(1+2 \ell)=-k\,.
\eea

When none of the exponent pairs differ by an integer, that is, when none of $\mathfrak{a},\mathfrak{b}, \mathfrak{c}-\mathfrak{a},\mathfrak{c}-\mathfrak{b}$ is an integer (requiring strictly $\omega\ne 0$), we have the following pairs of fundamental solutions
\bea
 \hat{R}_s(z)&=& (1-z)^p\,z^q\, (c_1 F[\mathfrak{a},\mathfrak{b},\mathfrak{c};z]+c_2 \,z^{1-\mathfrak{c}} \,F[\mathfrak{a}-\mathfrak{c}+1,\mathfrak{b}-\mathfrak{c}+1,2-\mathfrak{c};z])\,,\nonumber
\eea
%
The first term represents an ingoing wave $\hat{R}\sim z^{-i\mathcal{\alpha_+}}$ at the outer horizon $z=0$, while the second term represents an outgoing wave at the horizon. We are working at the classical level, so there can be no outgoing flux across the horizon, and thus we set $c_2=0$. Equivalently, we can now define the solution to the radial wave equation through (\ref{sols1}) yielding \footnote{The wave function for a scalar ($s=0$) was computed some time ago in e.g. \cite{Castro:2010fd} Equation (6.1)
\bea
\hat{R}&=& \left(\frac{r-r_+}{r-r_-}\right)^{-i \frac{2Mr_+}{r_+-r_-}(\omega-m\Omega)}(r-r_-)^{-1-\ell}\\
&&F\left(1+\ell-i \frac{4M}{r_+-r_-}(M\omega -r_+ m\Omega),1+\ell-2iM\omega,1-i\frac{4Mr_+}{r_+-r_-}(\omega-m\Omega);\frac{r-r_+}{r-r_-}\right)\,. \nonumber
\eea}.  
\begin{small}
\bea\label{eq:radialR}
\hat{R}&=& \left(\frac{r-r_+}{r-r_-}\right)^{-i \frac{2Mr_+}{r_+-r_-}(\omega-m\Omega)-s/2}(r-r_-)^{-1-\ell}\\
&&F\left(1+\ell-i \frac{4M(M\omega -r_+ m\Omega),}{r_+-r_-},1+\ell-2iM\omega-s,1-i\frac{4Mr_+}{r_+-r_-}(\omega-m\Omega)-s;\frac{r-r_+}{r-r_-}\right)\,. \nonumber
\eea 
\end{small}

One is now interested in the the behavior of the ingoing near-zone solution for large $r$ (or as $z\rightarrow 1$). Through the utilization of Kummer's solutions, the fundamental solutions can be explicitly connected into alternative expressions for hypergeometric functions relevant in $r\rightarrow \infty$ ($z\rightarrow 1$). For this computation step, there are two distinct ways in which to proceed. The coefficients $\ell$ are taken to be integers, without any additional frequency corrections. This will be our selected approach for the calculations in Section \ref{stricly}. A more refined approach can be used  where  the angular eigenvalues  $\ell$ are no longer assumed to be integers for non-zero frequencies and are accompanied by  corrections to the frequency involving $a \,\omega$ (see Appendix \ref{nonstricly}). It will be shown that both scenarios encompass the same aspects of the Love number coefficients, aligning exactly as well with the low-frequency findings presented in references like \cite{Mano:1996vt}.
%

\subsection{Tidal Response Coefficients for Kerr}
\label{stricly}

Starting from the solution to the radial equation \eqref{eq:radialR} in this section we compute the dynamical tidal response coefficients for Kerr. We consider the $K_{\ell,s}$ coefficients strictly as integers with no further $a \omega$ frequency corrections. This approach was employed in the context of the Kerr/CFT e.g.  \cite{Castro:2010fd}. While the method captures many essential features of the dynamical tidal deformations of the Kerr black hole, additional considerations will be required to fully describe these coefficients (see Appendix \ref{nonstricly}). It is important to highlight that $\mathfrak{c}=\mathfrak{a}+\mathfrak{b}-k$, where  $k$ is a positive integer, while $\mathfrak{c}$  itself is not an integer. This distinction allows us to derive the transformation of these solutions within the vicinity of the boundary point $z=1$ where the tidal response coefficients can be easily determined. The linear transformation applicable to the hypergeometric function (15.3.12 in Abramowitz \& Stegun) is as follows:
\bea\label{function}
F[\mathfrak{a},\mathfrak{b},\mathfrak{a}+\mathfrak{b}-k;z]&=&\frac{\Gamma(k) \Gamma(\mathfrak{a}+\mathfrak{b}-k)}{\Gamma(\mathfrak{a})\Gamma(\mathfrak{b})}\sum_{n=0}^{k-1}\frac{(\mathfrak{a}-k)_n(\mathfrak{b}-k)_n}{n!(1-k)_n}(1-z)^{(n-k)}\, \\
&-&\frac{(-1)^k\Gamma(\mathfrak{a}+\mathfrak{b}-k)}{\Gamma(\mathfrak{a}-k)\Gamma(\mathfrak{b}-k)}\sum_{n=0}^{\infty} \frac{(\mathfrak{a})_n(\mathfrak{b})_n}{n!(n+k)!}(1-z)^n [\log(1-z)+k_E]\,,\nonumber
\eea
where
\bea
k_E=\psi(\mathfrak{a}+n)+\psi(\mathfrak{b}+n)-\psi(n+1)-\psi(n+k+1)\,,
\eea
the $(x)_n\equiv\Gamma(x+n)/\Gamma(x)$ are the Pochammer symbols, and $\psi(x)\equiv {\Gamma'(x)}/{\Gamma(x)}$ the digamma functions. An important difference to the static cases studied in section \ref{sec:Review} above is that (\ref{function}) does not only consists on powers of $z$ but also contain logarithmic divergences.

We can now obtain the large $r \rightarrow \infty$ behavior of the ingoing wave solution by considering that $(1-z)\rightarrow (r_+-r_-)/r$ so that
\bea
&&R\xrightarrow[r\rightarrow\infty] \,\tilde{c_1} \left[\frac{\Gamma(k) \Gamma(\mathfrak{a}+\mathfrak{b}-k)}{\Gamma(\mathfrak{a})\Gamma(\mathfrak{b})}\, r^{\ell}+\frac{\Gamma(\mathfrak{a}+\mathfrak{b}-k)}{k!\,\Gamma(\mathfrak{a}-k)\Gamma(\mathfrak{b}-k)} \log\left(\frac{r_+-r_-}{r}\right)\, r^{-\ell-1}\right]
\eea
Here we can only record the coefficient of the logarithmic term in the ratio of fall-offs. provides us with the following dimensionless generic response coefficient
\bea
k_{\ell m}(\omega)=\frac{\Gamma(\mathfrak{a})\,\Gamma(\mathfrak{b})}{\Gamma(k+1)\,\Gamma(\mathfrak{a}-k)\Gamma(\mathfrak{b}-k)} \log\left(\frac{r_+-r_-}{r}\right)\,.
\eea
 in units of $( r_+-r_-)^{2\ell+1}$. After some simplifications we obtain 
\bea\label{lovenumberdyn}
k_{\ell m}(\omega)&=&\frac{\Gamma\left(1+\ell-i\frac{4 M}{r_+-r_-} (M\omega -r_+ m \Omega)\right)\,\Gamma(1+\ell- 2 i M \omega-s)}{(2\ell+1)!\,\Gamma(2\ell+1)\,\Gamma\left(-\ell-i\frac{4 M}{r_+-r_-} (M\omega -r_+ m \Omega)\right)\Gamma(-\ell- 2 i M \omega-s)} \nonumber\\
&&\times \left(\frac{r_+-r_-}{r_++r_-}\right)^{(1+2\ell)}\log\left(\frac{r_+-r_-}{r}\right)\nonumber\\
&=& \frac{4 M i (M\omega -r_+ m \Omega) ( 2 i M \omega+s)}{(2\ell+1)!\,\Gamma(2\ell+1) \, (r_+-r_-)} \left[\prod_{m=1}^{\ell} \left(m^2+\frac{16 M^2(M\omega -r_+ m \Omega)^2}{(r_+-r_-)^2} \right)\right]\nonumber\\
&&\times \left[\prod_{n=1}^{\ell} (n^2+( 2 M \omega- i s)^2 )\right]   \left(\frac{r_+-r_-}{r_++r_-}\right)^{(1+2\ell)} \log\left(\frac{r}{r_+-r_-}\right)  \,.
\eea
To obtain this expression we utilized the relation given in \eqref{relation3}. We argue that this expression corresponds to the frequency-dependent Love numbers for Kerr black holes. The coefficient in question depends on the distance at which the response is measured, represented by the parameter $r$, with $r_+-r_-$ serving as the renormalization scale. While the frequency-dependent part is only an approximation and may not fully capture the black hole response at non-zero frequencies, our systematic calculation shows that the response coefficients exhibit logarithmic contributions. This implies that the coefficients are universal and independent of the renormalization scheme. Any additional corrections not contained in the logarithmic part of the dynamical Love number \eqref{lovenumberdyn} are scheme-dependent. Consequently, we can conclude that the expression is valid for all orders in $\omega$ and will not receive any further contributions.

It is important to note that the coefficients \eqref{lovenumberdyn} in the Kerr black hole near-region are only valid for $\omega \neq 0$ or for frequency-dependent perturbations. The linear transformation for the hypergeometric function \eqref{function}  depends on $\mathfrak{b} \notin \mathbb{Z}$. However, in the limit of $\omega \to 0$, we observe that $\mathfrak{b} = \ell+ 1 - s \in \mathbb{Z}$, and the transformation involving logarithmic terms does not apply. In fact, for the static case, the response coefficients are given by \eqref{KerrLove2}. Therefore, it is necessary to consider the appropriate limits and conditions when using these expressions. We provide further remarks below.

\begin{itemize}

\item {\it No frequency-dependent dissipation in Kerr by scalar perturbations ($s=0$)}

The Love numbers $k_{\ell m}(\omega)$ in the near-region of a Kerr black hole are dynamic and all become real valued when the spin is zero $s=0$, as indicated by equation \eqref{lovenumberdyn}. The dynamical coefficient's imaginary response, represented by $\Im[k_{\ell m}(\omega)]=0$, is zero, indicating the absence of dissipation in the system, as shown by the vanishing of $\nu_{\ell}(\omega)=0$ for $\forall \omega$ for any frequency values. 
However this is not always the case, e.g. in the static case when $\omega=0$. It is important to note that Eq. \eqref{lovenumberdyn} is valid strictly for $\omega\ne 0$. As demonstrated by the static dissipative response for Kerr given in equation \eqref{KerrLove2} static dissipation is present for scalar fields $s=0$. This response indicates that there is some level of dissipation in the system under certain conditions, despite the initial absence of dissipation implied by the dynamical tidal response coefficients.

\item {\it The dynamical Love numbers for Kerr black holes and dissipative coefficients can be defined, in the slow-frequency limit, order by order in the frequency}

The response coefficient that we derived in our analysis, which is given by equation (\ref{lovenumberdyn}), can be further expanded in terms of the frequency of the external perturbation. This allows us to define the Love numbers order by order 
\bea\label{eq:dynamicalLove}
k_{lm}(\omega) 
=\sum_{n=1}^{\infty} k_{lm}^{(n)} \,\, \omega^n
=\sum_{n=1}^{\infty}(\kappa^{(n)}+ i \, \nu^{(n)}) \,\left(\frac{r_+-r_-}{r_++r_-}\right)^{(1+2l)}\, \omega^n \,,
\eea
where the first few orders yield
\bea
 \kappa^{(1)}&=&\frac{(-1)^{l+s+1} (l+s)!\,\Gamma(1+l-s)\,\Gamma\left(1+l+2 m i \gamma\right)}{(2l+1)!\,\Gamma(2l+1)\,\Gamma\left(-l+2 m i \gamma \right)} \log\left(\frac{r_+-r_-}{r}\right)2 M  i \,,\\
&=&\frac{(-1)^{s} (l+s)!\,(l-s)!}{(2l+1)!\,(2l)!} \ 4 M \,m\,\gamma\,  \log\left(\frac{r_+-r_-}{r}\right) \prod_{n=1}^l (n^2+ 4 m^2 \gamma^2) \,\\
&=& \nu^{(0)}\ 4 M \log\left(\frac{r_+-r_-}{r}\right)\,\\
 \kappa^{(2)}&=&  \kappa^{(1)}  \frac{4iM^2}{ (r_+-r_-)} (\psi(-l+2i m \gamma)-\psi(1+l+2i m \gamma)) \,,\\
 &=&- \kappa^{(1)}  \frac{4M^2}{ (r_+-r_-)} \left(\frac{1}{2m \gamma}+ 2 \sum^{l}_{n=1} \frac{2m \gamma}{(2m \gamma)^2+(n^2)}\right)\\
\nu^{(2)}&=&  \kappa^{(1)} \, 2 M (\psi(1+l+s)-\psi(1+l-s))\\
&=& \kappa^{(1)} \, 2 M \sum_{n=0}^{2 s-1}\frac{1}{n+\ell+1-s}\,.
\eea
and a new spin parameter
\bea
\gamma= a/(r_+-r_-)\,.
\eea
where $ \nu^{(0)}=\nu_{\ell m}(\omega=0)$ as defined in \eqref{KerrLove2}. In the last equations involving polygamma functions , we used several series representations that we derived in Appendix \ref{sec:gamma}.

These results show some interesting features of Kerr black hole solutions. On the one hand, in contrast to the static tidal response, the dynamical Love numbers $ \kappa^{(n>0)}$ for Kerr black holes do not vanish even at first order of the frequency $\omega$. Further, as one can easily observe from \eqref{eq:dynamicalLove}, the dynamical Love numbers for Kerr are generically not zero at all orders in the frequency $\omega$ and exhibit logarithmic running, in agreement with expectations of Wilsonian naturalness. On the other hand, to leading order in the frequency, the distortion coefficients $ k_{\ell m}^{(1)}\ne 0$ are real valued implying there is no dissipation  $\nu^{(1)}=0$ for Kerr black holes at this order \footnote{To calculate the Love numbers for a Kerr black hole in  the slow-frequency limit, we used the Taylor expansions of the Gamma function that correspond to natural values of the orbital integer number $\ell$. To obtain these expansions, we need to shift the argument of the expansion from $\ell$ to $\ell+\epsilon$, where $\epsilon \ll 1$, and consider $\omega \ll 1$, which results in a finite limit $1/\Gamma(-\ell- 2 M i \omega)=(-1)^{\ell+1} \ell! \, 2  M i \omega + \mathcal{O}(\omega^2)\,.$
Further, to achieve this, the gamma relation \eqref{relation3}, was employed, yielding the expression
\bea \label{approx2}
\frac{\Gamma(\ell+1+ 2 m i\gamma)}{\Gamma(-\ell+ 2 m i\gamma)}=(-1)^{\ell} \, 2  m i \gamma \prod_{n=1}^{\ell}(n^2+ 4 m^2 \gamma^2)\,.
\eea}. 

The expression  $k_{\ell m}^{(2)}=\kappa^{(2)}+ i \nu^{(2)}$ involves digamma functions. Certain combinations of digamma functions are either purely imaginary or real. Therefore we can identify the corresponding dynamical Love number and dissipative response. By closely examining the expression we draw two conclusions. First, for $s=0$, the dissipation coefficient vanishes $\nu^{(2)}=0$ so that there is no dissipation for scalar perturbations. Second, for $s=1,2$, the dissipation coefficients do not vanish at second order in the frequency $\omega^2$ .

\item {\it Kerr black holes do not universally behave like rigidly rotating dissipative spheres}

Expanding the response coefficient at the leading order in the black hole's spin is possible. We obtain the expression for the Love number coefficient at leading order in the frequency for slowly rotating Kerr black holes
\bea
\kappa_{\ell m}(\omega) =\frac{(-1)^{s} (\ell+s)!\,(\ell-s)\ell!^2}{(2\ell+1)!\,(2\ell)!} \  \, 4 M \omega \gamma  \,m \log\left(\frac{r_+-r_-}{r}\right) +\mathcal{O}(\gamma^2\omega)\,.
\eea
To leading order in the frequency and spin $\gamma$, the distortion coefficient exhibit only a purely real logarithmic correction. This suggests that dissipation, particularly in terms of imaginary contributions at this order, appears to be absent. Any additional contributions, which lack logarithmic behavior, can potentially vary based on the chosen scheme. Earlier works \cite{Charalambous:2021mea} suggested that Kerr black holes behaved at leading order in $\omega$ as rigidly rotating spheres due to non-logarithmic dissipative contributions. We argue that the existence of purely logarithmic corrections is proof that the prescribed dissipation in \cite{Charalambous:2021mea} is not universal.

We can directly compare this result with the time-dependent Love numbers in the low frequency regime obtained in the form of a series over hypergeometric functions, see Refs. \cite{Mano:1996vt,Sasaki:2003xr}. Further details will be presented in Section \ref{nonstricly}. The coefficients in black hole near-region approximation captures exactly the $\mathcal{O}(\omega \gamma)$ corrections to the tidal response coefficients (see equation $(6.16)$ in \cite{Charalambous:2021mea}). In the static limit $\omega \rightarrow0$, the Love number vanishes, matching previous results and the Newtonian expression in \cite{Charalambous:2021mea,Bonelli:2021uvf}.

\end{itemize}

\subsection{Response Coefficients for Schwarzschild}

Suppose now that  the black hole is not rotating. At a technical level the case of $a=0$ is straightforward. In order to apply this limit, we must specify $\Omega \rightarrow0$, $r_+\rightarrow 2 M$ and $r_- \rightarrow0$ in (\ref{lovenumberdyn}), so the dynamical tidal response coefficient for Schwarzschild is
\bea\label{lovenumberdynSch}
k^{Schw}_{\ell}(\omega)&=&\frac{\Gamma(1+\ell-s- 2 i M \omega)\Gamma(1+\ell- 2 i M \omega)}{(2\ell+1)!\,\Gamma(2\ell+1)\,\Gamma(-\ell-s- 2 i M \omega)\,\Gamma(-\ell- 2 i M \omega)} \log\left(\frac{2M}{r}\right)\,\\
&=& \frac{ ( 2 i M \omega  \,\, s- 4 M^2 \omega^2 )}{(2\ell+1)!\,\Gamma(2\ell+1) } \left[\prod_{j=1}^{\ell} \left(j^2+4 M^2 \omega^2 \right)\right] \left[\prod_{n=1}^{\ell} (n^2+( 2 M \omega- i s)^2 )\right] \log\left(\frac{r}{2M}\right)\,.\nonumber
\eea

\begin{itemize}
\item {\it Scalar perturbations do not dissipate}\\
 The dynamical Love numbers for Schwarzschild black holes with $s=0$ are
\bea
k^{Schw}_{\ell}(\omega)&=&\frac{\Gamma(1+\ell- 2 i M \omega)^2}{(2\ell+1)!\,\Gamma(2\ell+1)\,\Gamma(-\ell- 2 i M \omega)^2} \log\left(\frac{2M}{r}\right)\,\\
&=&-\frac{4 M^2  \omega^2 }{(2\ell+1)!\,\Gamma(2\ell+1)} \, \prod_{n=1}^l (n^2+ 2 M  \omega)^2 \log\left(\frac{2M}{r}\right)
\eea
The coefficients are purely real valued. We can therefore conclude that there is no dissipation for $s=0$ for all values of the frequency $\omega$.

\item {\it No dissipation in the slow-frequency regimes}\\
Let us analyze this expression in powers of the frequency $\omega$ by Taylor expanding (\ref{lovenumberdynSch}). Employing the low frequency expansion of the dynamical Love response (\ref{lovenumberdyn}) at leading order, one can show that the expression reduces to 
\bea
k^{Schw}_{\ell}(\omega) =  \frac{(-1)^{s+1}  \ell! (\ell+s)!\Gamma(1+\ell)\Gamma(1+\ell-s)}{(1+2\ell)! \Gamma(1+2\ell)} \,8 M^2 \omega^2 \log\left(\frac{2M}{r}\right)+\mathcal{O}(\omega^3)\,.
\eea

 There is no dissipation at first order of the frequency $\omega$ in Schwarzschild black holes.
 
\item {\it Agreement with post-Newtonian}\\
 As was stated, the Love numbers provide a measure of the tidal deformation of a non-rotating compact body  The Love numbers acquire
an operational meaning through the definition of tidally induced multipole moments. The black hole's response to an applied tidal field is measured by the Love numbers.
While the static coefficients are well known, the nonlinear number time-derivative numbers were not. A computation of the multipole moments was recently carried out
to the first post-Newtonian order by \cite{Poisson:2020vap}. Coincidently, the results of the tidal coefficients seem to agree with our results.
 For spin $s=2$ the expression reduces to
\bea \label{postnew}
k^{Schw}_{s=2}(\omega) =  -\frac{  (\ell-2)! (\ell-1)! \, \ell!(\ell+2)!}{2\,(1+2\ell)! (2\ell-1)!} \,4 M^2 \omega^2 \log\left(\frac{2M}{r}\right)+\mathcal{O}(\omega^3)\,.
\eea
After introducing a cutoff parameter, denoted as $r$, which restricts the range of integration, we find that our results are in perfect agreement with Eq. (1.3) in the reference \cite{Poisson:2020vap}. To address the logarithmic correction in Love numbers, we adopt the regularization scheme proposed in the PN limit \cite{Poisson:2020vap}. By choosing a characteristic length or energy scale of the system, such as $\log\left(\frac{2M}{r}\right) \sim 1$, we are able to resolve the logarithmic correction issue in Eq. \eqref{postnew}. It would be of great interest to conduct a more comprehensive comparison in future research, delving into the intricacies of the topic.

\item {\it In the low-frequency regime, the dynamical Love numbers for Schwarzschild black holes vanish for odd powers of $\omega^{2n+1}$ when $s=1,2$}\\
By closely examining the expression \eqref{lovenumberdynSch}, one can identify the dynamical Love numbers for Schwarzschild black holes, which are defined as the real part of $k_{\ell}(\omega)$. Our analysis reveals that in the slow-frequency regime, the dynamical Love numbers for Schwarzschild black holes vanish for odd powers of $\omega^{2n+1}$ due to bosons and graviton perturbations with $s=1$ or $2$.
 
\item {\it In the slow-frequency regime, there is frequency-dependent dissipation by Schwarzschild black holes only in odd powers of $\omega^{2n+1}$ when $s=1,2$}\\
In the slow-frequency limit, Schwarzschild black holes exhibit only a frequency-dependent dissipation in odd powers of the frequency $\omega^{2n+1}$ for $s=1,2$. 
 
\end{itemize}


\section{Comparison with Low Frequency Solutions}
\label{sec:Low}

Another method to solve the Teukolsky equations at finite frequency can be obtained in the form of  series over hypergeometric functions see \cite{Mano:1996vt}. Importantly, these coefficients are not purely imaginary  and can be determined analytically. A systematic calculation shows that Love numbers receive corrections already at linear order in $\omega$
\bea
k_{\ell m}(\omega) &=&
 \frac{\Gamma(-2\nu-1)\,\Gamma(1+\nu-s- 2 i M \omega)\,\Gamma\left(1+\nu+2i Q\right)}{\Gamma(2\nu+1)\,\Gamma(- \nu-s-2 i M \omega)\,\Gamma\left(- \nu+2i Q\right)} \, \left[1- 2 \Delta \ell \,\log\left(\frac{r_+-r_-}{r}\right)\right]\nonumber \\
 && \times \, (1+A_{\ell m} \,\,\omega) \left(\frac{r_+-r_-}{r_++r_-}\right)^{(1+2l)}+\mathcal{O}(\omega^2)
\eea
where 
\bea\label{eq:Qvalue}
Q=-\frac{2 M}{r_+-r_-} (M\omega -r_+ m \Omega) \,,
\eea 
The renormalized angular momentum is defined as $\nu \equiv \ell+\Delta \ell$, and $(1+A_{\ell m} \,\,\omega)$ are certain constant coefficients that are determined with a specific recursion relation \cite{Mano:1996vt}. 
After some simplifications, it can be shown that 
\bea
k_{\ell m}(\omega) = k^{(1)}_{\ell m} \, \omega +\mathcal{O} (\omega^2)\,,
\eea
with
\bea
k^{(1)}_{\ell m}=\frac{(-1)^{s} (\ell+s)!\,(\ell-s)!}{(2\ell+1)!\,(2\ell)!} \  \, 4 M \omega \gamma  \,m \log\left(\frac{r_+-r_-}{r}\right) \left(\prod_{n=1}^{\ell} (n^2+ 4 m^2 \gamma^2)\right)  \left(\frac{r_+-r_-}{r_++r_-}\right)^{(1+2 \ell)}  \,.
\eea
See Eq. (3.5) in \cite{Charalambous:2022rre}. The frequency-dependent part that we derive \eqref{lovenumberdyn} captures exactly Kerr's black hole response numbers at low-frequencies in $\omega$. This seems to indicate that the there is a technical advantage of the near-zone approach that we consider, in contrast with other proposals (such as Starobinsky \cite{Hui:2022vbh} or Love \cite{Charalambous:2021kcz} KEGs). It provides a simple frequency-dependent extension which allows to capture co-rotating modes while keeping the complexity of the solution the same as in the static case. 


\section{Love Numbers from Kerr Effective Geometries}
\label{sec:Effective}

Other near-horizon effective metrics such as the one we focused on for Kerr black holes \cite{Castro:2010fd} have been proposed to analyze Kerr's tidal response coefficients. In this context, these non-equivalent other cases were introduced as the near black hole zone approximations in \cite{Charalambous:2021kcz,Hui:2022vbh,Hui:2021vcv,Maldacena:1997ih}.
The primary objective of the approach is to construct a truncation of the full Teukolsky equation that can be exactly solved within a specific vicinity of the black hole. The approximation is designed to be accurate within a region that is sufficiently large to encompass both the vicinity of the black hole horizon (including the ergosphere) and an overlap with the asymptotically flat region.

The tidal response coefficients in the so-called Starobinsky KEG \cite{Charalambous:2021kcz} or Love KEG \cite{Hui:2022vbh} near-zone approximations are of the form
\bea\label{nearzone}
k^{Eff}_{\ell m}=\frac{\Gamma(-2\ell-1)\Gamma(1+\ell-s)\Gamma(1+\ell+2i\bar{Q})}{\Gamma(2\ell+1)\,\Gamma(- \ell-s)\,\Gamma(- \ell+2i\bar{Q})} \left( \frac{r_+-r_-}{r_++r_-}\right)^{(1+2\ell)}
\eea
where 
\bea
\bar{Q}= Q-M \omega
\eea
with $Q$ previously defined in \eqref{eq:Qvalue}. This expression for the Love numbers for Starobisnky KEG were first defined in equation (3.4) of \cite{Charalambous:2022rre}, and for Love KEG through Eq. (5.22) in \cite{Kehagias:2022ndy}.
The tidal response coefficients \eqref{nearzone} are imaginary, and represent dissipative effects with zero Love numbers (no real coefficient). It is important to remember though that the result  \eqref{nearzone} is approximate and does not capture the full black hole response at non-zero $\omega$. In fact, these coefficients do not correctly reproduce the low-frequency regime in \cite{Mano:1996vt}. In the low-frequency regime, as shown in \cite{Charalambous:2021kcz}, the coefficient is
\bea\label{nearzone}
k^{Eff}_{\ell m}\sim i r_s (\omega - m\Omega) (-1)^s \frac{ (\ell+ s)!(\ell- s)!(\ell!)^2}{ (2\ell)! (2\ell+ 1)!}+\mathcal{O}(\omega\Omega,\omega^2,\Omega^2)\,.
\eea
We can see that the KEGs near field approximation miss the $\mathcal{O}(\omega a)$ and $\mathcal{O}(\omega a \log(r_+-r_-/r))$ corrections. 
The $k^{Eff}_{\ell m}$ only capture the dissipative (imaginary) responses, but fail to capture the dynamical Love responses already at first order in $\omega$. In particular, the logarithmic invariant behavior present in \eqref{lovenumberdyn}. In contrast, the results obtained in our paper \eqref{lovenumberdyn}, with an alternative near-zone approximation (which is neither Starobinsky or Love KEG) does correctly reproduce the coefficients from the analysis in \cite{Mano:1996vt}. This seems to indicate that there is an advantage to employing the near-zone represented in \eqref{radialOmega} rather than other proposals \cite{Charalambous:2021kcz,Hui:2022vbh}.

\section{CFT Interpretation}
\label{sec:CFT}

Because the dynamical Love numbers \eqref{lovenumberdyn} measure the response to an incoming wave by the near region black hole, it is proportional to a two-point function in the CFT. Our modes have a boundary condition specified with no outgoing flux from horizon. We argue that the relevant two-point function is the retarded Green's function
\bea
G_R(\omega) \sim k_{\ell m}(\omega) \,.
\eea
To compare with the dual CFT we write $k_{\ell m}(\omega)$ in terms of the CFT temperatures $(T_L,T_R)=(\frac{r_++r_-}{4\pi a},\frac{r_+- r_-}{4\pi a})$, the linearization of their conjugate charges $(\omega_L,\omega_R)=(2 M^2\omega /a,\omega_L-m)$ and the conformal weights $(h_L,h_R)=(\ell,\ell)$ the dynamical Love number (\ref{lovenumberdyn}) becomes, for $s=0$
\bea
k_{\ell m}(\omega_R,\omega_L)&=&\frac{\Gamma\left(1+h_R-i\frac{\omega_R}{2\pi T_R}\right)\,\Gamma(1+h_L -i\frac{\omega_L}{2\pi T_L})}{(2\ell+1)!\,2\ell!\,\Gamma\left(-h_R-i\frac{\omega_R}{2\pi T_R}\right)\Gamma(-h_L- i \frac{\omega_L}{2\pi T_L})} \log\left(\frac{r_+-r_-}{r}\right)\\
&=&\sinh\left(\frac{\omega_R}{2T_R}\right)\sinh\left(\frac{\omega_L}{2T_L}\right)\left(h_R^2 +\left(\frac{\omega_R}{2\pi T_R}\right)^2\right) \left(h_L^2 +\left(\frac{\omega_L}{2\pi T_L}\right)^2\right) \\
&&\times \Bigg\vert{\Gamma\left(h_R-i\left(\frac{\omega_R}{2\pi T_R}\right)\right)}\Bigg\vert^2\ \Bigg\vert{\Gamma\left(h_L-i\left(\frac{\omega_L}{2\pi T_L}\right)\right)}\Bigg\vert^2\frac{\log\left(\frac{r}{r_+-r_-}\right)}
{(h_R+h_L+1)!(h_R+h_L)!}\nonumber     
\eea
where we used some identities found in Appendix {\ref{sec:gamma}}. The Love numbers are proportional to Green's function that in turn contains information about the physical excitations of the spacetime, the quasi-normal modes (QNM).  However, the consideration $ \omega M \ll 1$ which assumes real-valued frequencies $\omega$ prevents us from finding the imaginary QNM spectrum.
The agreement between the CFT and gravity results for small values of $\omega M$ relies on the specific values of the parameters involved. Consequently, the findings presented in this paper provide additional support to the notion that there exists a universal conformal symmetry governing the behavior of Kerr black holes, which encompasses the discussions found in references \cite{Chanson:2022wls} covering different aspects and specific scenarios. Our aim is to gain a more comprehensive understanding of this matter in the future.

\section{Discussion}
\label{sec:Discussion}

In this work we computed the dynamical Love numbers for Kerr black holes. Our systematic calculation, consistent with a CFT description of the Kerr black hole, and manifestly preserving $SL(2,R) \times SL(2,R)$ hidden symmetries, shows that Love numbers receive logarithmic corrections at all orders in the frequency $\omega$. 
\bea
\kappa^{Kerr}_{\ell m}(\omega\ne0)\ne0\,,\qquad \nu^{Kerr}_{\ell m}(\omega\ne0)\ne0
\eea
The presence of time-dependent tidal deformability represents a dynamical coupling between the black hole and its exterior environment. This seems to indicate that spinning Kerr black holes deform under an external stationary gravitational field. Our results relied on an important technique. The introduction of the effective Kerr near-zone with an $SL(2,R) \times SL(2,R)$ hidden symmetry. There is a technical advantage, with other proposals such as Starobinsky or Love KEGs. It provides a simple frequency-dependent extension which allows to capture co-rotating modes while keeping the complexity of the solution the same as in the static case. Another confirmation comes from the direct comparison with the slow-frequency solutions \cite{Mano:1996vt} where we find a complete agreement. 

Non-vanishing dynamical Love numbers are in stark contrast with past studies related to Love numbers for Kerr in static external gravitational fields where these vanish $\kappa^{Kerr}_{\ell m}(\omega=0)=0$. This seems to create some tension with the so called Love symmetry results in e.g. \cite{Hui:2022vbh,Charalambous:2021mea} which argue that the corresponding Love numbers vanish for all frequencies $\omega$. The near-field approximations, with either $SL(2,R) \times U(1)$ hidden symmetry in \cite{Charalambous:2021kcz} or involving an $SO(4,2)$ hidden symmetry in the work \cite{Hui:2022vbh}, do not capture the low frequency $\mathcal{O}(\omega)$ and logarithmic behavior of the dynamical Love numbers. 

For an extremal Kerr black holes ($r_+=r_-$) the analysis on the dynamical Love numbers does not apply. The wave-equation equation considered \eqref{radial} develops an irregular singularity at $r=r_+=r_-$ in the extremal limit. Nevertheless, there exists a consistent near zone truncation of the Teukolsky equation even in the extremal limit where this computation of the dynamical tidal response for Kerr black holes can be solved. We leave this question for future work.

There is one results that may bring also insight on time dependent tidal deformations of Schwarzschild black holes (in the limit of $a \rightarrow 0$). Our calculations, equation \eqref{postnew}, reveal a precise match  (after fixing the characteristic length) between the tidal deformations and the post-Newtonian (PN) computations documented in reference \cite{Poisson:2020vap}. This is particularly relevant in the context of binary inspirals, where the resulting gravitational waveform exhibits sensitivity to the Love numbers, becoming noticeable starting at the 5PN order in the Post-Newtonian expansion (refer to citations  \cite{Vines:2011ud,Bini:2012gu,Flanagan:2007ix}). 

Notably, the imaginary contribution in \eqref{lovenumberdyn} primarily introduces dissipation. In the case of binary inspirals, dissipative response coefficient become significant at the 2.5PN order, effectively capturing the spin exchange between the binary components -- as detailed in \cite{Poisson:1994yf,Alvi:2001mx,Poisson:2004cw,Porto:2007qi,Tagoshi:1997jy,Goldberger:2005cd,Chatziioannou:2012gq}. Our results also reveal a discernible pattern: all the tidal coefficients are directly proportional to the static coefficient. This pattern appears to be explicable by the hidden symmetries within the solutions and could potentially manifest itself in the QNM spectrum, waveform characteristics and photon rings. Previous studies, as indicated in \cite{Charalambous:2021mea}, suggested that Kerr black holes exhibit rigidly rotating behavior at the leading order in $\omega$. This behavior was attributed to non-logarithmic dissipative factors. We argue that the existence of purely logarithmic corrections in our results serves as evidence that rotating black holes behave as non-rigidly rotational objects.

Our research can naturally be extended in a number of directions. We could, for instance, we can investigate the tidal deformability in the context of spinning black holes: considering various scenarios such as high-frequency regimes, black holes with significant rotation, or within the context of strong tidal fields. Another intriguing direction would be to extend our methodology to higher dimensions, specifically investigating whether the hidden symmetry of $SL(2,R)\times SL(2,R)$ plays a role in Myers-Perry black holes. Furthermore, we can inquire whether the dynamical Love numbers derived within this framework exhibits a symmetry akin to the one proposed in  \cite{Chanson:2022wls,Rodriguez:2023xjd}.

Other extensions of our results can be directly constructed simply upon inserting the corresponding expressions for the outer and inner horizons $r_{\pm}$ in the static tidal coefficients for black holes \bea\label{lovenumberdynKerrNUT0}
k_{\ell m}(\omega=0)=  \frac{\Gamma(-2\ell -1)\, \Gamma(1+\ell -s)\, \Gamma(1+\ell+2mi\gamma)}{\Gamma(- \ell-s)\,\Gamma(- \ell+2 m i \gamma)\,\Gamma(2\ell+1)} \left( \frac{r_+-r_-}{r_++r_-}\right)^{(1+2\ell)}\,,
\eea
and the dynamical expressions
\bea\label{lovenumberdynKerrNUT}
k_{\ell m}(\omega)&=&\frac{\Gamma\left(1+\ell-i\frac{4 M}{r_+-r_-} (M\omega -r_+ m \Omega)\right)\,\Gamma(1+\ell- 2 i M \omega-s)}{(2\ell+1)!\,\Gamma(2\ell+1)\,\Gamma\left(-\ell-i\frac{4 M}{r_+-r_-} (M\omega -r_+ m \Omega)\right)\Gamma(-\ell- 2 i M \omega-s)} \nonumber\\
&&\times \left(\frac{r_+-r_-}{r_++r_-}\right)^{(1+2\ell)}\log\left(\frac{r_+-r_-}{r}\right)
\eea
where we defined $\gamma= a/(r_+-r_-)$. Other parameters in the solution are linked to the mass $M$ and spin parameter $a$.
In particular, if we consider the replacement of the horizon expression in \eqref{lovenumberdynKerrNUT0} and \eqref{lovenumberdynKerrNUT} for black hole solutions exhibiting an $SL(2,R)\times SL(2,R)$ hidden symmetry as in the Kerr-NUT black holes
\bea
r_{\pm}\rightarrow r^{KTN}_{\pm}=M\pm\sqrt{M^2+N^2-a^2}\,,
\eea
and the so called Kerr-MOG black hole of the Scalar Tensor Vector Gravity (STVG), also known as modified gravity (MOG) theory \cite{Guo:2018kis} 
\bea
r_{\pm}\rightarrow r^{MOG}_{\pm}=r(1+\alpha)\pm\sqrt{M^2(1+\alpha)-a^2}\,,
\eea
The parameter $\alpha$ is the deformation parameter, such that when $\alpha=0$ one recovers the Kerr black hole solution.
We find that the static tidal coefficients \eqref{lovenumberdynKerrNUT0} for both, Kerr-NUT and Kerr-MOG black holes, are imaginary, and we therefore deduce that the Love number vanish. These observations for Kerr-NUT black holes can be an opening a window for further physical interpretation of these solutions and their potential astrophysical applications. For example, possible NUT signatures in microlensing data have been considered \cite{lynden1998classical,bogdanov2008search}. We also observe, for Kerr Taub-NUT and Kerr-MOG black holes, that dynamical Love number are finite.


\section*{Acknowledgements}

We thank Luca Santoni for helpful discussions. We would like to thank the Mitchell Family Foundation at the Cook's Branch workshop and the Centro de Ciencias de Benasque Pedro Pascual for the hospitality where in 2023 some of the research was carried out. The work of MP is supported by an STFC consolidated grant ST/L000415/1, String Theory, Gauge Theory and Duality. MJR is partially supported through the NSF grant PHY-2012036, PHY-2309270, RYC-2016-21159, CEX2020-001007-S and PGC2018-095976-B-C21, funded by MCIN/AEI/10.13039/501100011033. 

\appendix

\section{Gamma Function}
\label{sec:gamma}

The Gamma function is defined by the integral formula,
\bea
\Gamma(z)=\int_{0}^{\infty} t^{z-1} e^{-t} dt\,.
\eea
The integral converges absolutely for $Re(z)>0$. For integer values, $n \ge 0$, it is related to the factotial
\bea
\Gamma(n+1)= n!\,,
\eea
As an extension to complex numbers $z$, the function satisfies the functional equations
\bea
\Gamma(z+1)= z \,\Gamma(z)\,.
\eea
and
\bea
\Gamma(z)\Gamma(1-z)  = \frac{\pi}{\sin \pi z}.
\eea

Two other important relations employed in this paper include
\\

{\bf Relation 1:}\\
\bea \label{relation3}
\frac{\Gamma(l+1+ i A)}{\Gamma(-l+ i A)}=(-1)^{l} \, i A \prod_{n=1}^l (n^2+A^2)\,.
\eea
{\it Proof:}\\
\\
We show here the different steps in the derivation
\bea
\frac{\Gamma(l+1+ i A)}{\Gamma(-l+ i A)}&&= \frac{\Gamma(l+1+ i A)\,\Gamma(l+1- i A) }{\Gamma(-l+ i A)\,\Gamma(l+1- i A) }\\
&&=\frac{\pi A}{{\Gamma(-l+ i A)\,\Gamma(l+1- i A) }\sinh(\pi A)} \prod_{n=1}^l (n^2+A^2)
\eea
where we used $|\Gamma(1+l+iB)|^2=\frac{\pi B}{\sinh(\pi B)} \prod_{n=1}^l (n^2+B^2)$. Note that employing the relation $\Gamma(z)\Gamma(1-z)=1/\sin(\pi z)$ we find
\bea
\frac{\Gamma(l+1+ i A)}{\Gamma(-l+ i A)}&&=\frac{\pi A \sin(\pi (l+1- i A))}{\sinh(\pi A)} \prod_{n=1}^l (n^2+A^2)\\
&&=\frac{\pi A (-1)^l i \sinh(A \pi)}{\sinh(\pi A)} \prod_{n=1}^l (n^2+A^2)\\
&&=(-1)^{l} \, i A \prod_{n=1}^l (n^2+A^2)\,.
\eea
In the first line we employed the relation $\sin(\pi(l+1- i A))=(-1)^l i \sinh(A)$.
\\

{\bf Relation 2:}\\
\\
The difference between digamma functions $\psi$ yields
\bea \label{relation4}
\psi(m+i A)-\psi(n+iA)=-i\left(\frac{1}{A}+ 2 \sum^{l}_{\bar{q}=1} \frac{A}{A^2+(\bar{q}^2)}\right)
\eea

{\it Proof:}
\bea
\psi(m+i A)-\psi(n+iA)&=&\left. \frac{d}{dz} \log\left( \Gamma(z+i A)\right)\right|_{z=m}-\left.\frac{d}{dz} \log\left( \Gamma(z+i A)\right)\right|_{z=n}\\
&=&\frac{d}{dz} \log\left( \frac{\Gamma(z+i A)}{\Gamma(z-p+i A)}\right)\\
&=&\sum^{p-1}_{q=0}\frac{1}{(n+q+i A)}\\
&=&\sum^{2l}_{q=0}\frac{1}{(-l+q+i A)}\\
&=&\sum^{l}_{\bar{q}=-l}\frac{1}{(\bar{q}+i A)}\\
&=&\frac{1}{iA}+\sum^{l}_{\bar{q}=1}\left(\frac{1}{\bar{q}+iA}+\frac{1}{-\bar{q}+iA}\right)\\
&=&\frac{1}{i A}+ 2 \sum^{l}_{\bar{q}=1} \frac{i A}{(iA)^2-\bar{q}^2}
\eea
were we considered $m=l+1$, $n=-l$, $m-n=p=2l+1$, and $\bar{q}=q-l$.


\section{Alternative Calculation of the Dynamical Response Coefficient for Kerr Black Holes}
\label{nonstricly}

In this section we compute the dissipative response coefficient for Kerr black holes in the $\omega M \ll 1$ regime. The corresponding radial Teukolsky equation reduces to \eqref{radial}. So far we have considered the leading contribution $\mathcal{O}(1)$ to separation $K_{\ell,s}$ coefficients \eqref{separationconst} in a small frequency expansion $a\omega$. But the angular eigenvalues receive further corrections for small $a\omega$ given schematically by \eqref{separationconst} to order $\mathcal{O} (a^2\omega^2)$ becoming non-integer real valued eigenvalue. Further, this expression provides a rational behind the the analytic continuation $\ell \in \mathbb{R}$. To include the higher order corrections an analytic continuation can be used. The analytic continuation has been employed in several works on Love numbers \cite{Kol:2011vg} to avoid the ambiguity of the source/response splitting in the definitions of the tidal coefficients. This is effectively equivalent to promoting the orbital spin number $\ell$ to non-integer values. 

The starting point to implement the proposed analytical continuation is the solution to the radial wave equation (\ref{sols1}) yielding ingoing boundary conditions on the horizon. The scalar tidal response coefficients can be extracted by means of the analytic continuation at $r \rightarrow \infty$, or equivalently $z\rightarrow 1$ behavior of the ingoing near-region solution. To achieve this aim one uses the $z \rightarrow 1-z$ transformation law for the hypergeometric function
\bea
F[\mathfrak{a},\mathfrak{b},\mathfrak{c};z]&=&\,\frac{\Gamma(\mathfrak{c})\Gamma(\mathfrak{c}-\mathfrak{a}-\mathfrak{b})}{\Gamma(\mathfrak{c}-\mathfrak{a})\Gamma(\mathfrak{c}-\mathfrak{b})}F[\mathfrak{a},\mathfrak{b},\mathfrak{a}+\mathfrak{b}-\mathfrak{c}+1;1-z]\nonumber\\
 &&+z^{\mathfrak{c}-\mathfrak{a}-\mathfrak{b}}\,\frac{\Gamma(\mathfrak{c})\Gamma(\mathfrak{a}+\mathfrak{b}-\mathfrak{c})}{\Gamma(\mathfrak{a})\Gamma(\mathfrak{b})} F[\mathfrak{c}-\mathfrak{a},\mathfrak{c}-\mathfrak{b},\mathfrak{c}-\mathfrak{a}-\mathfrak{b}+1;1-z]\,.\nonumber
\eea
and the property $F[\mathfrak{a},\mathfrak{b},\mathfrak{c};0]=1$.
Taking all these behaviors at $r \rightarrow \infty$ we can identify the Love numbers from the decaying mode 
\bea\label{eq:expansion}
\hat{R}_s(r) \xrightarrow[r\rightarrow\infty] \, \tilde{c}_1 \, r^{\ell} \left(1+\left( \frac{r}{r_+-r_-}\right)^{-(1+2 \ell)} \frac{\Gamma(\mathfrak{a})\Gamma(\mathfrak{b})\Gamma(\mathfrak{c}-\mathfrak{b}-\mathfrak{a})}{\Gamma(\mathfrak{c}-\mathfrak{a})\Gamma(\mathfrak{c}-\mathfrak{b})\Gamma(\mathfrak{a}+\mathfrak{b}-\mathfrak{c})}\right)\,,
\eea
which comparing with the expression in (\ref{perturbation}) provides us with the following dimensionless response coefficients
\bea\label{dissipativedyn}
k_{\ell m}(\omega)&=&\frac{\Gamma(\mathfrak{a})\Gamma(\mathfrak{b})\Gamma(\mathfrak{c}-\mathfrak{b}-\mathfrak{a})}{\Gamma(\mathfrak{c}-\mathfrak{a})\Gamma(\mathfrak{c}-\mathfrak{b})\Gamma(\mathfrak{a}+\mathfrak{b}-\mathfrak{c})} \left( \frac{r_+-r_-}{r_++r_-}\right)^{(1+2\ell)}\,,\nonumber\\
&=& \frac{\Gamma(-2\ell-1)\,\Gamma(1+\ell-s- 2 i M \omega)\,\Gamma\left(1+\ell+2i Q\right)}{\Gamma(2\ell+1)\,\Gamma(- \ell-s-2 i M \omega)\,\Gamma\left(- \ell+2i Q\right)} \left( \frac{r_+-r_-}{r_++r_-}\right)^{(1+2 \ell)}\,.
\eea
where 
\bea
Q=-\frac{2 M}{r_+-r_-} (M\omega -r_+ m \Omega)\,.
\eea
 To estimate the frequency dependent corrections $\Delta \ell$ we can perturb the tidal response coefficient \eqref{dissipativedyn} replacing $\ell \rightarrow \nu$ with the renormalized angular momentum 
\bea
\nu \equiv \ell+\Delta \ell=\ell+ a \omega \frac{2 m s^2}{\ell (\ell+1)(2 \ell+1)}+...
\eea
Importantly, the relevant response coefficients are not purely real in this case. The presence of a simple pole at $\epsilon=2M \omega$ is important. To see this we expand to linear order in $2M \omega$ while keeping all powers of $Q$ while also expanding the renormalized angular momentum where $\ell$ is an integer $\ell \ge |s|$ and $\Delta \ell = \mathcal{O} (2M \omega)$ in
\eqref{eq:expansion}. We find 
\bea
 \frac{\Gamma(-2\nu-1)}{\Gamma(2\nu+1)}&=&\frac{1}{2\Delta \ell (2 \ell+1)!(2 \ell)!}+\mathcal{O}(\epsilon^0)\\ 
\frac{\Gamma(1+\nu-s- 2 i M \omega)}{\Gamma(-\nu-s-2 i M \omega)} &=&(-1)^{\ell+s+1}(i\epsilon+\Delta \ell) (\ell-s)! (\ell+s)!+\mathcal{O}(\epsilon^2)\\ 
\frac{\Gamma(1+\nu+2iQ)}{\Gamma (-\nu+2iQ)} &=&(-1)^{\ell} \sin(2i Q \pi - \pi \Delta \ell) \frac{2 i Q}{\sin(2i Q \pi)} \prod_{n=1}^{\ell}(n^2+4 Q^2)+\mathcal{O}(\epsilon^2)
\eea
as well as a logarithm that comes from the Taylor expansion of $x^{- 2 \nu-1}$ such that
\bea
r^{-1- 2 \nu}= r^{-1- 2 \ell} (1-2 \Delta \ell \log r)
\eea
Extracting all the contributions from these coefficients we arrive at a finite logarithmic contribution
\bea
k_{ \ell m}(\omega)&=& \frac{(-1)^{s} (\ell+s)!\,(\ell-s)!}{(2\ell+1)!\,(2\ell)!} \  \, 4 M \omega \gamma  \,m \log\left(\frac{r_+-r_-}{r}\right) \prod_{n=1}^{\ell} (n^2+ 4 m^2 \gamma^2) \,\\
\eea
After some simplifications, it can be shown that 
\bea
k_{\ell m}(\omega) = k^{(1)}_{\ell m} \, \omega +\mathcal{O} (\omega^2)\,,
\eea
with
\bea
k^{(1)}_{\ell m}=\frac{(-1)^{s} (\ell +s)!\,(\ell -s)!}{(2\ell +1)!\,(2\ell )!} \  \, 4 M \omega \gamma  \,m \log\left(\frac{r_+-r_-}{r}\right) \left(\prod_{n=1}^{\ell} (n^2+ 4 m^2 \gamma^2)\right)  \left(\frac{r_+-r_-}{r_++r_-}\right)^{(1+2 \ell)}  \,.
\eea

\bibliographystyle{jheppub}
\bibliography{lovebib}

\providecommand{\href}[2]{#2}\begingroup\raggedright\begin{thebibliography}{10}

\bibitem{Gurlebeck:2015xpa}
N.~G\"urlebeck, ``{No-hair theorem for Black Holes in Astrophysical
  Environments},'' \href{http://dx.doi.org/10.1103/PhysRevLett.114.151102}{{\em
  Phys. Rev. Lett.} {\bfseries 114} no.~15, (2015) 151102},
  \href{http://arxiv.org/abs/1503.03240}{{\ttfamily arXiv:1503.03240 [gr-qc]}}.

\bibitem{Chatziioannou:2012gq}
K.~Chatziioannou, E.~Poisson, and N.~Yunes, ``{Tidal heating and torquing of a
  Kerr black hole to next-to-leading order in the tidal coupling},''
  \href{http://dx.doi.org/10.1103/PhysRevD.87.044022}{{\em Phys. Rev. D}
  {\bfseries 87} no.~4, (2013) 044022},
  \href{http://arxiv.org/abs/1211.1686}{{\ttfamily arXiv:1211.1686 [gr-qc]}}.

\bibitem{Barros_2022}
S.~C. C.~B. et~al., ``{Detection of the tidal deformation of WASP-103b at 3
  sigma with CHEOPS},'' {\em Astronomy and Astrophysics} {\bfseries 657} (2022)
  A52.

\bibitem{Binnington:2009bb}
T.~Binnington and E.~Poisson, ``Relativistic theory of tidal love numbers,''
  \href{http://dx.doi.org/10.1103/PhysRevD.80.084018}{{\em Phys. Rev. D}
  {\bfseries 80} (2009) 084018},
  \href{http://arxiv.org/abs/0906.1664}{{\ttfamily arXiv:0906.1664 [gr-qc]}}.

\bibitem{Fang:2005qq}
H.~Fang and G.~Lovelace, ``Tidal coupling of a schwarzschild black hole and
  circularly orbiting moon,''
  \href{http://dx.doi.org/10.1103/PhysRevD.72.124016}{{\em Phys. Rev. D}
  {\bfseries 72} (2005) 124016},
  \href{http://arxiv.org/abs/gr-qc/0505156}{{\ttfamily arXiv:gr-qc/0505156
  [gr-qc]}}.

\bibitem{Damour:2009vw}
T.~Damour and A.~Nagar, ``Relativistic tidal properties of neutron stars,''
  \href{http://dx.doi.org/10.1103/PhysRevD.80.084035}{{\em Phys. Rev. D}
  {\bfseries 80} (2009) 084035},
  \href{http://arxiv.org/abs/0906.0096}{{\ttfamily arXiv:0906.0096 [gr-qc]}}.

\bibitem{Kol:2011vg}
B.~Kol and M.~Smolkin, ``Black hole stereotyping: Induced gravito-static
  polarization,'' \href{http://dx.doi.org/10.1007/JHEP02(2012)010}{{\em JHEP}
  {\bfseries 02} (2012) 010}, \href{http://arxiv.org/abs/1110.3764}{{\ttfamily
  arXiv:1110.3764 [hep-th]}}.

\bibitem{Hui:2020xxx}
R.~P. L.~S. L.~Hui, A.~Joyce and A.~R. Solomon, ``Static response and love
  numbers of schwarzschild black holes,''
  \href{http://dx.doi.org/10.1088/1475-7516/2021/04/052}{{\em JCAP} {\bfseries
  04} (2021) 052}, \href{http://arxiv.org/abs/2010.00593}{{\ttfamily
  arXiv:2010.00593 [hep-th]}}.

\bibitem{Chia:2020yla}
H.~S. Chia, ``Tidal deformation and dissipation of rotating black holes,''
  \href{http://dx.doi.org/10.1103/PhysRevD.104.024013}{{\em Phys. Rev. D}
  {\bfseries 104} no.~2, (2021) 024013},
  \href{http://arxiv.org/abs/2010.07300}{{\ttfamily arXiv:2010.07300 [gr-qc]}}.

\bibitem{Goldberger:2020fot}
J.~L. W.~D.~Goldberger and I.~Z. Rothstein, ``Non-conservative effects on
  spinning black holes from world-line effective field theory,''
  \href{http://dx.doi.org/10.1007/JHEP06(2021)053}{{\em JHEP} {\bfseries 06}
  (2021) 053}, \href{http://arxiv.org/abs/2012.14869}{{\ttfamily
  arXiv:2012.14869 [hep-th]}}.

\bibitem{Charalambous:2021mea}
S.~D. P.~Charalambous and M.~M. Ivanov, ``On the vanishing of love numbers for
  kerr black holes,'' \href{http://dx.doi.org/10.1007/JHEP05(2021)038}{{\em
  JHEP} {\bfseries 05} (2021) 038},
  \href{http://arxiv.org/abs/2102.08917}{{\ttfamily arXiv:2102.08917
  [hep-th]}}.

\bibitem{Goldberger:2004jt}
W.~D. Goldberger and I.~Z. Rothstein, ``An effective field theory of gravity
  for extended objects,''
  \href{http://dx.doi.org/10.1103/PhysRevD.73.104029}{{\em Phys. Rev. D}
  {\bfseries 73} (2006) 104029},
  \href{http://arxiv.org/abs/hep-th/0409156}{{\ttfamily arXiv:hep-th/0409156
  [hep-th]}}.

\bibitem{Porto:2016pyg}
R.~A. Porto, ``The effective field theorist's approach to gravitational
  dynamics,'' \href{http://dx.doi.org/10.1016/j.physrep.2016.04.003}{{\em Phys.
  Rept.} {\bfseries 633} (2016) 1--104},
  \href{http://arxiv.org/abs/1601.04914}{{\ttfamily arXiv:1601.04914
  [hep-th]}}.

\bibitem{Vines:2011ud}
E.~E.~F. J.~Vines and T.~Hinderer, ``Post-1-newtonian tidal effects in the
  gravitational waveform from binary inspirals,''
  \href{http://dx.doi.org/10.1103/PhysRevD.83.084051}{{\em Phys. Rev. D}
  {\bfseries 83} (2011) 084051},
  \href{http://arxiv.org/abs/1101.1673}{{\ttfamily arXiv:1101.1673 [gr-qc]}}.

\bibitem{Bini:2012gu}
T.~D. D.~Bini and G.~Faye, ``Effective action approach to higher-order
  relativistic tidal interactions in binary systems and their effective one
  body description,'' \href{http://dx.doi.org/10.1103/PhysRevD.85.124034}{{\em
  Phys. Rev. D} {\bfseries 85} (2012) 124034},
  \href{http://arxiv.org/abs/1202.3565}{{\ttfamily arXiv:1202.3565 [gr-qc]}}.

\bibitem{Poisson:1994yf}
E.~Poisson and M.~Sasaki, ``Gravitational radiation from a particle in circular
  orbit around a black hole. 5: Black hole absorption and tail corrections,''
  \href{http://dx.doi.org/10.1103/PhysRevD.51.5753}{{\em Phys. Rev. D}
  {\bfseries 51} (1995) 5753--5767},
  \href{http://arxiv.org/abs/gr-qc/9412027}{{\ttfamily arXiv:gr-qc/9412027
  [gr-qc]}}.

\bibitem{Alvi:2001mx}
K.~Alvi, ``Energy and angular momentum flow into a black hole in a binary,''
  \href{http://dx.doi.org/10.1103/PhysRevD.64.104020}{{\em Phys. Rev. D}
  {\bfseries 64} (2001) 104020},
  \href{http://arxiv.org/abs/gr-qc/0107080}{{\ttfamily arXiv:gr-qc/0107080
  [gr-qc]}}.

\bibitem{Poisson:2004cw}
E.~Poisson, ``Absorption of mass and angular momentum by a black hole:
  Time-domain formalisms for gravitational perturbations, and the small-hole /
  slow-motion approximation,''
  \href{http://dx.doi.org/10.1103/PhysRevD.70.084044}{{\em Phys. Rev. D}
  {\bfseries 70} (2004) 084044},
  \href{http://arxiv.org/abs/gr-qc/0407050}{{\ttfamily arXiv:gr-qc/0407050
  [gr-qc]}}.

\bibitem{Porto:2007qi}
R.~A. Porto, ``Absorption effects due to spin in the worldline approach to
  black hole dynamics,''
  \href{http://dx.doi.org/10.1103/PhysRevD.77.064026}{{\em Phys. Rev. D}
  {\bfseries 77} (2008) 064026},
  \href{http://arxiv.org/abs/0710.5150}{{\ttfamily arXiv:0710.5150 [hep-th]}}.

\bibitem{Lowe:2011aa}
D.~A. Lowe and A.~Skanata, ``Generalized hidden kerr/cft,''
  \href{http://dx.doi.org/10.1088/1751-8113/45/47/475401}{{\em J. Phys. A}
  {\bfseries 45} (2012) 475401},
  \href{http://arxiv.org/abs/1112.1431}{{\ttfamily arXiv:1112.1431 [hep-th]}}.

\bibitem{Charalambous:2021kcz}
S.~D. P.~Charalambous and M.~M. Ivanov, ``Hidden symmetry of vanishing love
  numbers,'' \href{http://dx.doi.org/10.1103/PhysRevLett.127.101101}{{\em Phys.
  Rev. Lett.} {\bfseries 127} no.~10, (2021) 101101},
  \href{http://arxiv.org/abs/2103.01234}{{\ttfamily arXiv:2103.01234
  [hep-th]}}.

\bibitem{Hui:2022vbh}
L.~Hui, A.~Joyce, R.~Penco, L.~Santoni, and A.~R. Solomon, ``{Near-zone
  symmetries of Kerr black holes},''
  \href{http://dx.doi.org/10.1007/JHEP09(2022)049}{{\em JHEP} {\bfseries 09}
  (2022) 049}, \href{http://arxiv.org/abs/2203.08832}{{\ttfamily
  arXiv:2203.08832 [hep-th]}}.

\bibitem{Castro:2010fd}
A.~M. A.~Castro and A.~Strominger, ``Hidden conformal symmetry of the kerr
  black hole,'' {\em Phys. Rev. D} {\bfseries 82} (2010) 024008,
  \href{http://arxiv.org/abs/1004.0996}{{\ttfamily arXiv:1004.0996 [hep-th]}}.

\bibitem{Charalambous:2022rre}
P.~Charalambous, S.~Dubovsky, and M.~M. Ivanov, ``{Love symmetry},''
  \href{http://dx.doi.org/10.1007/JHEP10(2022)175}{{\em JHEP} {\bfseries 10}
  (2022) 175}, \href{http://arxiv.org/abs/2209.02091}{{\ttfamily
  arXiv:2209.02091 [hep-th]}}.

\bibitem{Mano:1996vt}
H.~S. S.~Mano and E.~Takasugi, ``Analytic solutions of the teukolsky equation
  and their low frequency expansions,''
  \href{http://dx.doi.org/10.1143/PTP.95.1079}{{\em Prog. Theor. Phys.}
  {\bfseries 95} (1996) 1079--1096},
  \href{http://arxiv.org/abs/gr-qc/9603020}{{\ttfamily arXiv:gr-qc/9603020
  [gr-qc]}}.

\bibitem{Poisson:2020vap}
E.~Poisson, ``Compact body in a tidal environment: New types of relativistic
  love numbers, and a post-newtonian operational definition for tidally induced
  multipole moments,''
  \href{http://dx.doi.org/10.1103/PhysRevD.103.064023}{{\em Phys. Rev. D}
  {\bfseries 103} no.~6, (2021) 064023},
  \href{http://arxiv.org/abs/2012.10184}{{\ttfamily arXiv:2012.10184 [gr-qc]}}.

\bibitem{Teukolsky:1973ha}
S.~A. Teukolsky, ``Perturbations of a rotating black hole. 1. fundamental
  equations for gravitational electromagnetic and neutrino field
  perturbations,'' \href{http://dx.doi.org/10.1086/152444}{{\em Astrophys. J.}
  {\bfseries 185} (1973) 635--647}.

\bibitem{Teukolsky:1972}
S.~A. Teukolsky, ``{Rotating black holes - separable wave equations for
  gravitational and electromagnetic perturbations},''
  \href{http://dx.doi.org/10.1103/PhysRevLett.29.1114}{{\em Phys. Rev. Lett.}
  {\bfseries 29} (1972) 1114--1118}.

\bibitem{Castro:2013kea}
A.~Castro, J.~M. Lapan, A.~Maloney, and M.~J. Rodriguez, ``{Black Hole
  Monodromy and Conformal Field Theory},''
  \href{http://dx.doi.org/10.1103/PhysRevD.88.044003}{{\em Phys. Rev. D}
  {\bfseries 88} (2013) 044003},
  \href{http://arxiv.org/abs/1303.0759}{{\ttfamily arXiv:1303.0759 [hep-th]}}.

\bibitem{LeTiec:2020spy}
A.~L. Tiec and M.~Casals, ``Spinning black holes fall in love,''
  \href{http://dx.doi.org/10.1103/PhysRevLett.126.131102}{{\em Phys. Rev.
  Lett.} {\bfseries 126} no.~13, (2021) 131102},
  \href{http://arxiv.org/abs/2007.00214}{{\ttfamily arXiv:2007.00214 [gr-qc]}}.

\bibitem{AB1964}
M.~{Abramowitz} and I.~A. {Stegun}, {\em Handbook of Mathematical Functions
  with Formulas, Graphs, and Mathematical Tables}.
\newblock Dover, New York City, ninth dover printing, tenth gpo printing~ed.,
  1964.

\bibitem{Castro:2013lba}
A.~M. A.~Castro, J. M.~Lapan and M.~J. Rodriguez, ``Black hole scattering from
  monodromy,'' \href{http://dx.doi.org/10.1088/0264-9381/30/16/165005}{{\em
  Class. Quant. Grav.} {\bfseries 30} no.~16, (2013) 165005},
  \href{http://arxiv.org/abs/1304.3781}{{\ttfamily arXiv:1304.3781 [hep-th]}}.

\bibitem{Hui:2021vcv}
R.~P. L.~S. L.~Hui, A.~Joyce and A.~R. Solomon, ``Ladder symmetries of black
  holes. implications for love numbers and no-hair theorems,''
  \href{http://dx.doi.org/10.1088/1475-7516/2022/01/032}{{\em JCAP} {\bfseries
  01} (2022) 032}, \href{http://arxiv.org/abs/2105.01069}{{\ttfamily
  arXiv:2105.01069 [hep-th]}}.

\bibitem{Sasaki:2003xr}
M.~Sasaki and H.~Tagoshi, ``{Analytic black hole perturbation approach to
  gravitational radiation},'' \href{http://dx.doi.org/10.12942/lrr-2003-6}{{\em
  Living Rev. Rel.} {\bfseries 6} (2003) 6},
  \href{http://arxiv.org/abs/gr-qc/0306120}{{\ttfamily arXiv:gr-qc/0306120}}.

\bibitem{Bonelli:2021uvf}
G.~Bonelli, C.~Iossa, D.~P. Lichtig, and A.~Tanzini, ``{Exact solution of Kerr
  black hole perturbations via CFT2 and instanton counting: Greybody factor,
  quasinormal modes, and Love numbers},''
  \href{http://dx.doi.org/10.1103/PhysRevD.105.044047}{{\em Phys. Rev. D}
  {\bfseries 105} no.~4, (2022) 044047},
  \href{http://arxiv.org/abs/2105.04483}{{\ttfamily arXiv:2105.04483
  [hep-th]}}.

\bibitem{Maldacena:1997ih}
J.~M. Maldacena and A.~Strominger, ``Universal low-energy dynamics for rotating
  black holes,'' \href{http://dx.doi.org/10.1103/PhysRevD.56.4975}{{\em Phys.
  Rev. D} {\bfseries 56} (1997) 4975--4983},
  \href{http://arxiv.org/abs/hep-th/9702015}{{\ttfamily arXiv:hep-th/9702015
  [hep-th]}}.

\bibitem{Kehagias:2022ndy}
D.~P. A.~Kehagias and A.~Riotto, ``Quasinormal modes and love numbers of kerr
  black holes from ads$_{2}$ black holes,''
  \href{http://dx.doi.org/10.1088/1475-7516/2023/01/035}{{\em JCAP} {\bfseries
  01} (2023) 035}, \href{http://arxiv.org/abs/2211.02384}{{\ttfamily
  arXiv:2211.02384 [hep-th]}}.

\bibitem{Chanson:2022wls}
A.~Chanson, V.~Martin, M.~J. Rodriguez, and L.~F. Temoche, ``{CFT duals for
  black rings and black strings},''
  \href{http://dx.doi.org/10.1007/JHEP04(2023)066}{{\em JHEP} {\bfseries 04}
  (2023) 066}, \href{http://arxiv.org/abs/2212.12537}{{\ttfamily
  arXiv:2212.12537 [hep-th]}}.

\bibitem{Flanagan:2007ix}
E.~E. Flanagan and T.~Hinderer, ``Constraining neutron star tidal love numbers
  with gravitational wave detectors,''
  \href{http://dx.doi.org/10.1103/PhysRevD.77.021502}{{\em Phys. Rev. D}
  {\bfseries 77} (2008) 021502},
  \href{http://arxiv.org/abs/0709.1915}{{\ttfamily arXiv:0709.1915 [gr-qc]}}.

\bibitem{Tagoshi:1997jy}
H.~Tagoshi, S.~Mano, and E.~Takasugi, ``{PostNewtonian expansion of
  gravitational waves from a particle in circular orbits around a rotating
  black hole: Effects of black hole absorption},''
  \href{http://dx.doi.org/10.1143/PTP.98.829}{{\em Prog. Theor. Phys.}
  {\bfseries 98} (1997) 829--850},
  \href{http://arxiv.org/abs/gr-qc/9711072}{{\ttfamily arXiv:gr-qc/9711072}}.

\bibitem{Goldberger:2005cd}
W.~D. Goldberger and I.~Z. Rothstein, ``{Dissipative effects in the worldline
  approach to black hole dynamics},''
  \href{http://dx.doi.org/10.1103/PhysRevD.73.104030}{{\em Phys. Rev. D}
  {\bfseries 73} (2006) 104030},
  \href{http://arxiv.org/abs/hep-th/0511133}{{\ttfamily arXiv:hep-th/0511133}}.

\bibitem{Rodriguez:2023xjd}
A.~R.~S. M.~J.~Rodriguez, L.~Santoni and L.~F. Temoche, ``Love numbers for
  rotating black holes in higher dimensions,''
  \href{http://arxiv.org/abs/2304.03743}{{\ttfamily arXiv:2304.03743
  [hep-th]}}.

\bibitem{Guo:2018kis}
M.~Guo, N.~A. Obers, and H.~Yan, ``{Observational signatures of near-extremal
  Kerr-like black holes in a modified gravity theory at the Event Horizon
  Telescope},'' \href{http://dx.doi.org/10.1103/PhysRevD.98.084063}{{\em Phys.
  Rev. D} {\bfseries 98} no.~8, (2018) 084063},
  \href{http://arxiv.org/abs/1806.05249}{{\ttfamily arXiv:1806.05249 [gr-qc]}}.

\bibitem{lynden1998classical}
D.~Lynden-Bell and M.~Nouri-Zonoz, ``Classical monopoles: Newton, nut space,
  gravomagnetic lensing, and atomic spectra,'' {\em Reviews of Modern Physics}
  {\bfseries 70} no.~2, (1998) 427.

\bibitem{bogdanov2008search}
M.~Bogdanov and A.~Cherepashchuk, ``Search for exotic matter from gravitational
  microlensing observations of stars,'' {\em Astrophysics and Space Science}
  {\bfseries 317} no.~3-4, (2008) 181--192.

\end{thebibliography}\endgroup

\end{document}